\documentclass[aps,apl,reprint,superscriptaddress,longbibliography]{revtex4-2}

\usepackage{graphicx}
\usepackage{amsmath,amsfonts,amssymb}	
\usepackage{mathrsfs}
\usepackage[normalem]{ulem}
\usepackage[dvipsnames]{xcolor}
\usepackage{epstopdf}
\usepackage[retain-explicit-plus]{siunitx}
\usepackage[version=4]{mhchem}
\usepackage{graphicx}
\usepackage{tabularx}
\newcolumntype{Y}{>{\centering\arraybackslash}X}
\usepackage[section]{placeins}
\usepackage{longtable}
\usepackage{braket}
\usepackage{booktabs}
\usepackage{hyperref}

\hypersetup{
    colorlinks=true,
    linkcolor=blue,
    filecolor=magenta,      
    urlcolor=blue,
    citecolor=blue
    }
\definecolor{alizarin}{rgb}{0.82, 0.1, 0.26} 
\definecolor{violet}{rgb}{0.5, 0.1, 1} 
\definecolor{darkgreen}{rgb}{0, 0.8, 0} 

\draft 

\begin{document}

\title{AI-enhanced discovery and accelerated synthesis of metal phosphosulfides}


\author{Javier Sanz Rodrigo}
\affiliation{National Centre for Nano Fabrication and Characterization (DTU Nanolab), Technical University of Denmark, 2800 Kongens Lyngby, Denmark}

 \author{Nicholas A. Kryger-Nelson}
\affiliation{National Centre for Nano Fabrication and Characterization (DTU Nanolab), Technical University of Denmark, 2800 Kongens Lyngby, Denmark}

\author{Lena A. Mittmann}%
\affiliation{National Centre for Nano Fabrication and Characterization (DTU Nanolab), Technical University of Denmark, 2800 Kongens Lyngby, Denmark}

\author{Eugène Bertin}%
\affiliation{National Centre for Nano Fabrication and Characterization (DTU Nanolab), Technical University of Denmark, 2800 Kongens Lyngby, Denmark}

\author{Ivano E. Castelli}
\affiliation{Department of Energy Conversion and Storage, Technical University of Denmark, 2800 Kongens Lyngby, Denmark}

\author{Andrea Crovetto}%
\email{ancro@dtu.dk}
\affiliation{National Centre for Nano Fabrication and Characterization (DTU Nanolab), Technical University of Denmark, 2800 Kongens Lyngby, Denmark}


\begin{abstract}
Metal phosphosulfides have emerged as unique multifunctional materials, but they present unique synthesis challenges compared to more established material classes such as oxides and nitrides. As a consequence, experimental development and theoretical understanding of phosphosulfides have focused on individual compounds rather than on accelerated broad-range exploration.
In this work, we first evaluate the synthesizability and band gaps of 909 hypothetical ternary phosphosulfides by density functional theory.
We find 19 previously unknown thermodynamically stable compounds, including the first Si- and Ge-based phosphosulfides. For rapid band gap prediction, we then develop a multi-fidelity machine learning model to translate semilocal density functional theory band gaps into experimentally calibrated band gaps.
Importantly, we extend the accelerated material development workflow to the experimental domain by demonstrating a route to high-throughput synthesis and characterization of virtually any phosphosulfide material system.
The method is based on thin-film combinatorial libraries and yields over 100 unique compositions in each experiment, enabling us to synthesize four distinct phosphosulfide compounds in only four combinatorial experiments without prior synthesis recipes and without compromising on material quality.
Thus, we argue that accelerated materials development workflows combining theory, artificial intelligence, synthesis, and characterization can be viable even for experimentally challenging inorganic materials. 
\end{abstract}

\maketitle


\section{Introduction}

Metal phosphosulfides are an inherently versatile class of materials. As solid solutions between phosphides and sulfides, they allow for gradual property tuning to achieve optimal performance in a variety of applications~\cite{Kibsgaard2014,Park2014a}. As ordered compounds, there are over 200 experimentally reported phosphosulfides with unique properties that can't be simply interpolated from the corresponding sulfides and phosphides~\cite{mittmannPhosphosulfideSemiconductorsOptoelectronics2024}. 
The wide range of phosphorus oxidation states that are accessible in the presence of a more electronegative and a more electropositive species give a plethora of possibilities for materials design.

For example, Li-based phosphosulfides derived from \ce{Li10GeP2S12} are among the fastest ionic conductors at room temperature and natural candidates as solid electrolytes in solid-state batteries~\cite{bouguernComparativeAdvancesSulfide}. \ce{CuInP2S6} is an exotic ferroelectric material with record bulk photovoltaic effect performance~\cite{liEnhancedBulkPhotovoltaic2021}. Atomically-thin, antiferromagnetic \ce{NiPS3} exhibits novel spin-orbit-entangled excitons~\cite{Kang2020}. \ce{SnP2S6} possesses a unique combination of strong second-harmonic generation, resistance to laser damage, and phase matchability for a variety of applications in nonlinear optics~\cite{heSnP2S6PromisingInfrared2022}.

Despite their rich property portfolio, development of phosphosulfides has mainly progressed in the form of application-driven "material silos", in the sense that discovery of materials with interesting properties has typically been followed by searches for compositionally or structurally similar compounds. Examples are studies on Na-based phosphosulfides as battery materials following the discovery of Li-based superionic conductors~\cite{schnaubeltImpuritiesNa2SPrecursor2025}, on van der Waals phosphosulfides following the exfoliation of the first 2D phosphosulfides~\cite{yangUniversalStrategySynthesis2024}, and on non-centrosymmetric phosphosulfides as non-linear optical materials following the discovery of optically active P-S structural motifs~\cite{kongChalcophosphatesPlentifulSource2025}. The materials screening study by Han and Ebert~\cite{Han2021}, although broader in scope, only considered 18 ternary metal phosphosulfides.

\begin{figure*}[]
    \centering
    \includegraphics[width=1\textwidth]{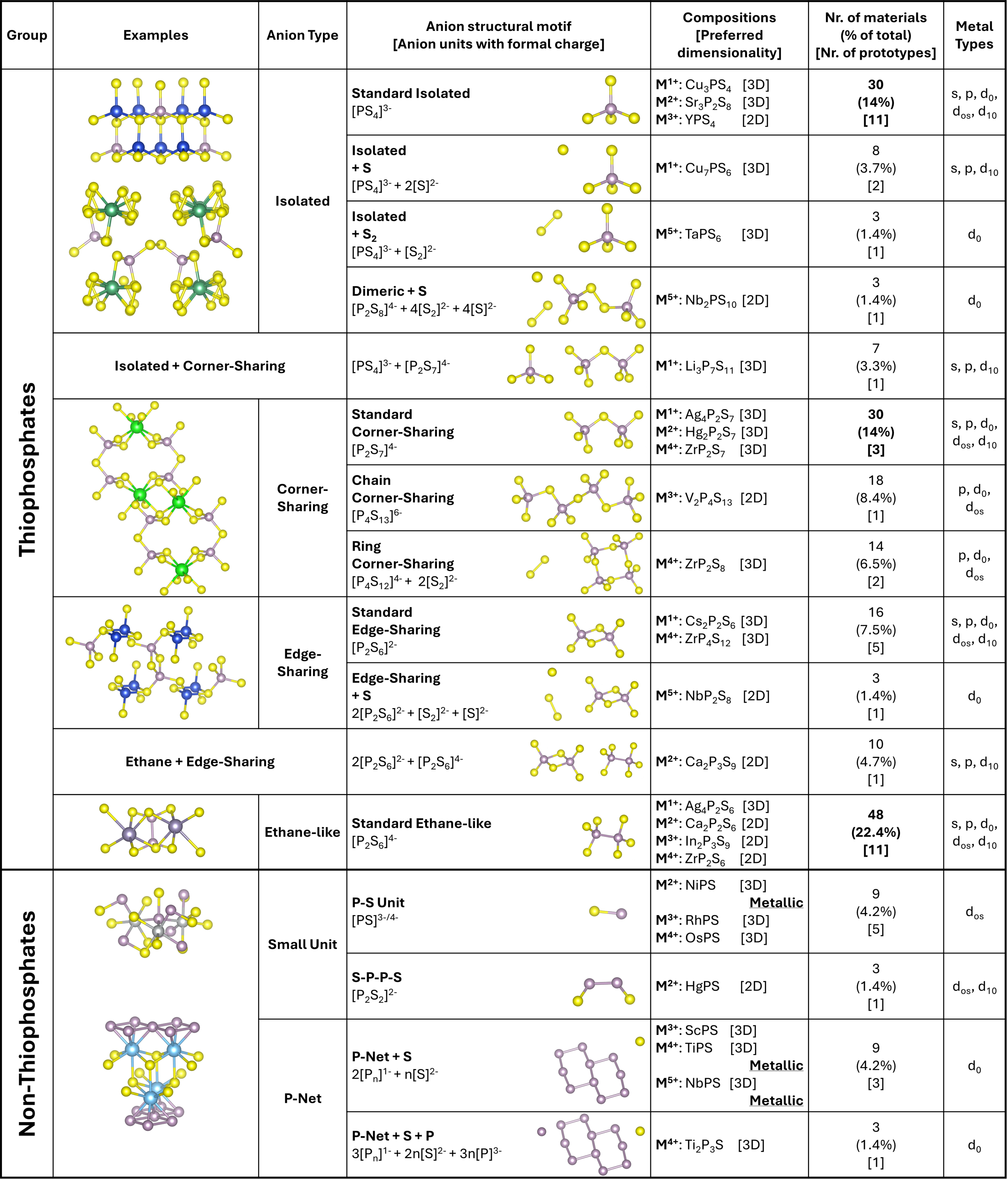}
    \caption{Summary of the structural and compositional diversity of phosphosulfides. The structures are visualized with VESTA~\cite{Momma2011}. Assignment of oxidation states is discussed in the SI. The "number of materials" (and the corresponding "\% of total") refers to the number of phosphosulfides calculated in this study that were found to be both lowest-energy polymorphs and within the \SI{100}{meV/atom} stability tolerance defined in the main text. When these numbers are in bold font, they indicate the three most common structural motifs. A full list of prototypes is given in Table~\ref{tab:mp-id}. When a composition is labeled as "metallic", it means that it is not charge balanced. The "metal types" are the blocks of the periodic table containing metals that are compatible with the corresponding structural motif. Among d-block metals, d$_0$, d$_{10}$, and d$_\mathrm{os}$ refer to the electronic configuration of the metals after compound formation, with d$_\mathrm{os}$ standing for open-shell configurations.
    \label{fig:motifs}}
\end{figure*}

We address two outstanding questions in this paper. First, how do application-agnostic properties of phosphosulfides (synthesizability and band gap) vary with composition and structure? We provide some answers based on the calculation of thermodynamic stability and band gaps of 909 ternary phosphosulfides through a combination of density functional theory (DFT) and multi-fidelity machine learning (ML). Along the way, we discover 19 previously unreported ternary phosphosulfides predicted to be thermodynamically stable.

All previously reported synthesis procedures for phosphosulfides known to us are serial, low-throughput processes. Even in papers that report many different compounds~\cite{yangUniversalStrategySynthesis2024,zhouCompositionPhaseEngineering2022,Byvik1982}, the materials were synthesized by one-material-at-a-time approaches that lack the ability to screen a range of material compositions and/or process parameters in parallel. Hence, the second question we address is: Can experimental development of phosphosulfides be accelerated by modern high-throughput methods, despite various challenges in dealing with phosphorus and sulfur in experimental settings? More broadly, this question is of interest for any class of materials where volatile, corrosive, and/or toxic synthesis precursors are necessary. Such a definition arguably includes most inorganic material families that are not oxides, nitrides or intermetallics. By realizing a previously presented vision for a high-throughput experimental setup compatible with "difficult" chemistries~\cite{mittmannPhosphosulfideSemiconductorsOptoelectronics2024}, we demonstrate rapid synthesis of four phosphosulfide compounds with high crystalline quality in just a few combinatorial experiments.

\section{Results}
\subsection{Common compositions and anion structures}

We classify all the phosphosulfides calculated in this study as either thiophosphates or non-thiophosphates (Fig.~\ref{fig:motifs}). In thiophosphates, phosphorus is present in a high positive oxidation state and bonds to sulfur but not to the metal M. This behavior is analogous to phosphates, hence the name. Phosphorus in thiophosphates can be seen either as a metal cation or as part of a polyanionic P-S motif with an overall negative oxidation state. In this paper we adopt the latter approach and visualize the many different thiophosphate polyanion geometries and formal oxidation states in Fig.~\ref{fig:motifs}. 
In non-thiophosphate compounds, phosphorus takes a negative or neutral oxidation state and bonds to the metal. Phosphorus may also bond to sulfur and to other phosphorus atoms, depending on the specific non-thiophosphate structure (see Fig.~\ref{fig:motifs}). 

\begin{figure} [h!]
        \centering
        \includegraphics[width=\columnwidth]{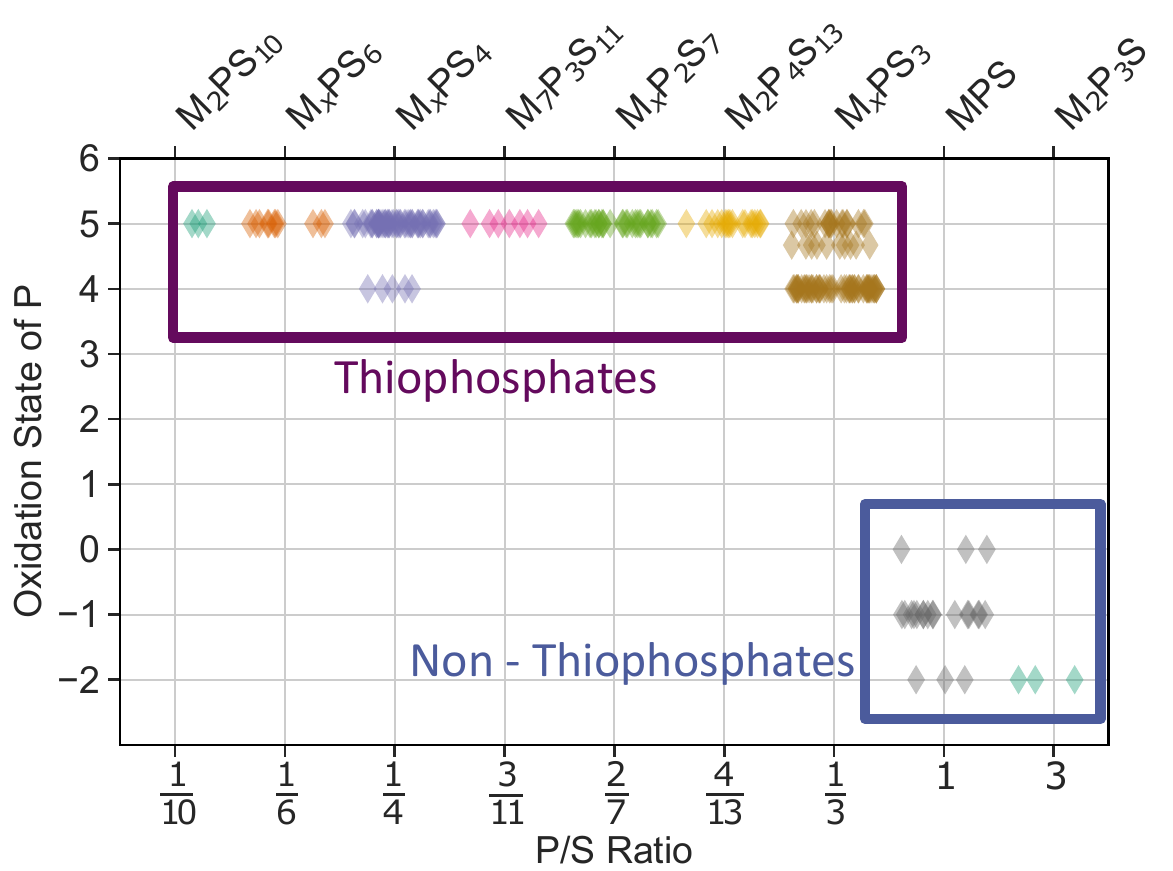}
        \caption{Oxidation state of phosphorus as a function of the atomic P/S ratio in the ternary phosphosulfides calculated in this study (only lowest-energy polymorphs within stability tolerance). The P/S ratios are discrete and are linked to the generalized compositions shown on the top x axis. The data points are jittered for a more intuitive visualization of the number of materials at each P/S ratio.}
        \label{fig:PtoSvsPOX}
\end{figure}

Thiophosphates are by far the dominant class of known phosphosulfides, as they make up 85\% of the experimentally synthesized ternary phosphosulfides (72 in total). We find that the transition between thiophosphates and non-thophosphates is well described by the atomic P/S ratio in the material (Fig.~\ref{fig:PtoSvsPOX}). When $ \mathrm{P/S} \leq \frac{1}{3}$, phosphorus is in the +5 or +4 oxidation state and the material is a thiophosphate. When $ \mathrm{P/S} > \frac{1}{3}$, phosphorus switches to oxidation states of -2, -1, and 0 (Fig.~\ref{fig:PtoSvsPOX}), marking the transition from thiophosphates to non-thiophosphates. Note that phosphorus is never present in the -3 oxidation state typical of III-V semiconductors, because even when P bonds to M in non-thiophosphates, it also bonds to either S or other P atoms~\cite{Crovetto2022b}.

Fig.~\ref{fig:motifs} summarizes the structural chemistry, elemental compositions, dimensionality, and occurrence of phosphosulfide anions found in at least a few different ternary phosphosulfides close to the stability hull. The structural chemistry of thiophosphates was also partially discussed in previous work~\cite{yangReviewStructuralDiversity2021,kongChalcophosphatesPlentifulSource2025}.
The basic building block of thiophosphate polyanions is the \ce{PS4} tetrahedron. When this basic unit is not connected to other polyanionic motifs, we refer to it as the "isolated" \ce{[PS4]^3-} polyanion, one of the most common polyanions encountered in thiophosphate compounds. The isolated \ce{PS4} tetrahedron can be found together with \ce{[S2]^2-} dimers or with individual \ce{[S]^2-} anions or both, giving rise to materials with \ce{M_xPS6} and \ce{M2PS10} compositions.

In the "corner-sharing" family of thiophosphates (Fig.~\ref{fig:motifs}) \ce{PS4} tetrahedra are connected by sharing corners. The arrangement of corner-sharing tetrahedra can simply be two units (a \ce{[P2S7]^4-} polyanion, the third most common type in phosphosulfides), a linear chain, or a ring. These three arrangements are present in materials with compositions \ce{M_xP2S7} and \ce{M2P4S13}, and \ce{M4P2S8}, respectively. Materials with \ce{M7P3S11} composition feature both isolated \ce{PS4} tetrahedra and corner-sharing \ce{PS4} tetrahedra pairs in equal amounts.

In the "edge-sharing" family, a pair of \ce{PS4} tetrahedra is connected by sharing an edge (a \ce{[P2S6]^2-} polyanion), giving rise to materials with \ce{M_xP2S6} composition. When the edge-sharing motif is present alongside \ce{[S2]^2-} dimers and individual \ce{[S]^2-} anions, the materials have \ce{MP2S8} composition.

Interestingly, the "ethane-like" polyanion is the most common in thiophosphates (Fig.~\ref{fig:motifs}) despite being the only thiophosphate polyanion that is not based on \ce{PS4} tetrahedra. Instead, two trigonal pyramidal \ce{PS3} units are linked by a P-P bond, similar to the geometry of the ethane molecule. It is compatible with metals with all oxidation states up to +4, resulting in materials with \ce{M_xP2S6} compositions. Ethane-like and edge-sharing polyanions can simultaneously be present in equal amounts in the same materials, giving rise to \ce{M2P3S9} compositions.

Among non-thiophosphates, three anionic motifs for the \ce{MPS} composition are found: the P-S unit, made of a single P-S bond which is bonded to metal atoms from both ends; the S-P-P-S motif, made of two P-S units bonded to each other through a P-P bond; and the phosphorus net + sulfide anion motif, in which the metal bonds to a plane of P atoms and to individual sulfide anions.
Interestingly, the S-P-P-S motif with its bonded metal atoms is structurally very similar to the ethane-like polyanion, with the trigonal pyramidal \ce{PS3} units of the ethane-like polyanion replaced by trigonal pyramidal \ce{PM2S} units.

\begin{figure}[ht!]
    \centering
    \includegraphics[width=\columnwidth]{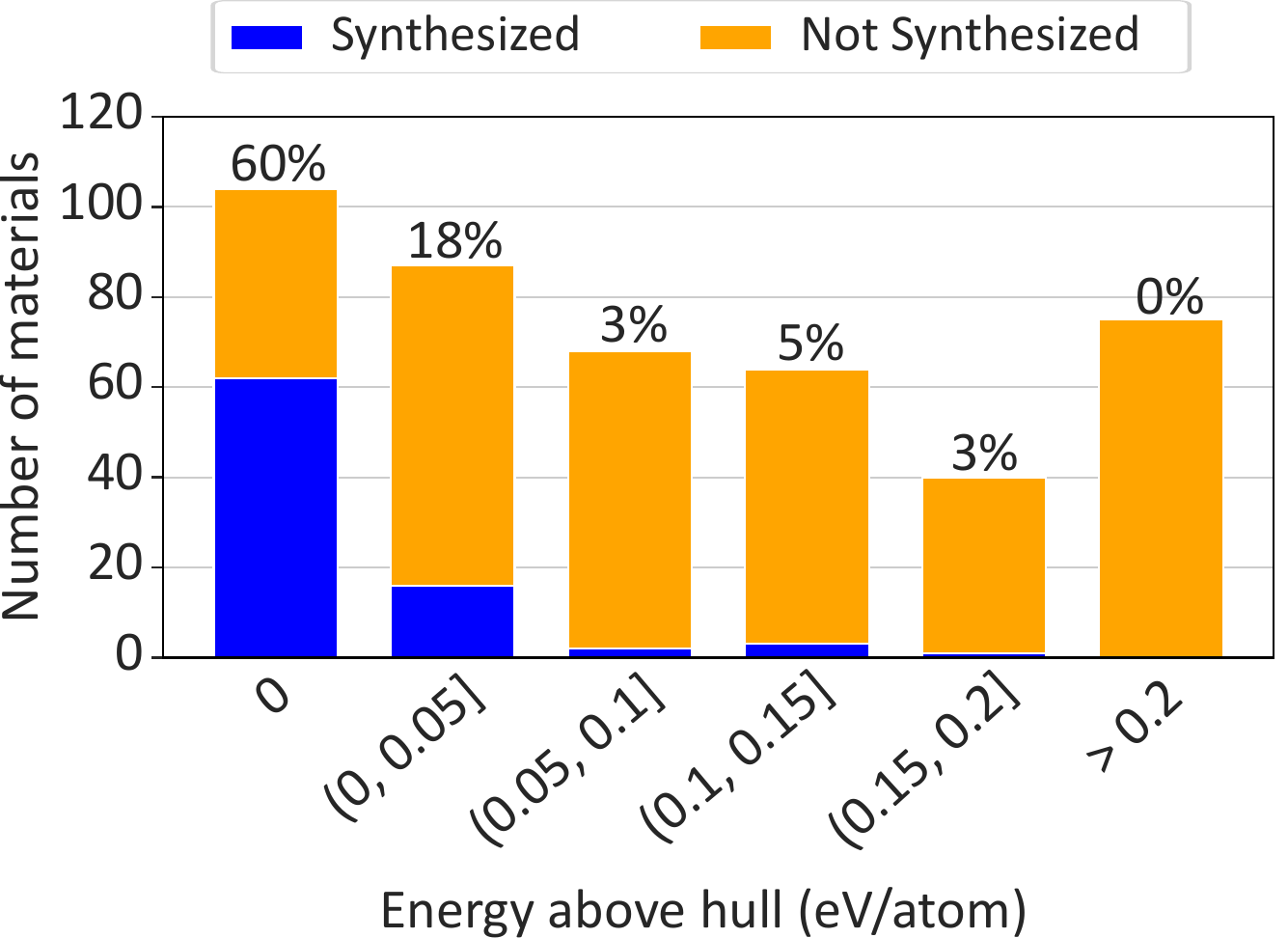}
    \caption{Number of synthesized phosphosulfides as a function of their energy above the stability hull. Only the lowest-energy polymorphs found in study are included in the statistics. The percentage of synthesized materials for each $E_h$ range is indicated.}
    \label{fig:Synth}
\end{figure}

\begin{figure*}[ht!]
\centering%
\includegraphics[width=\textwidth]{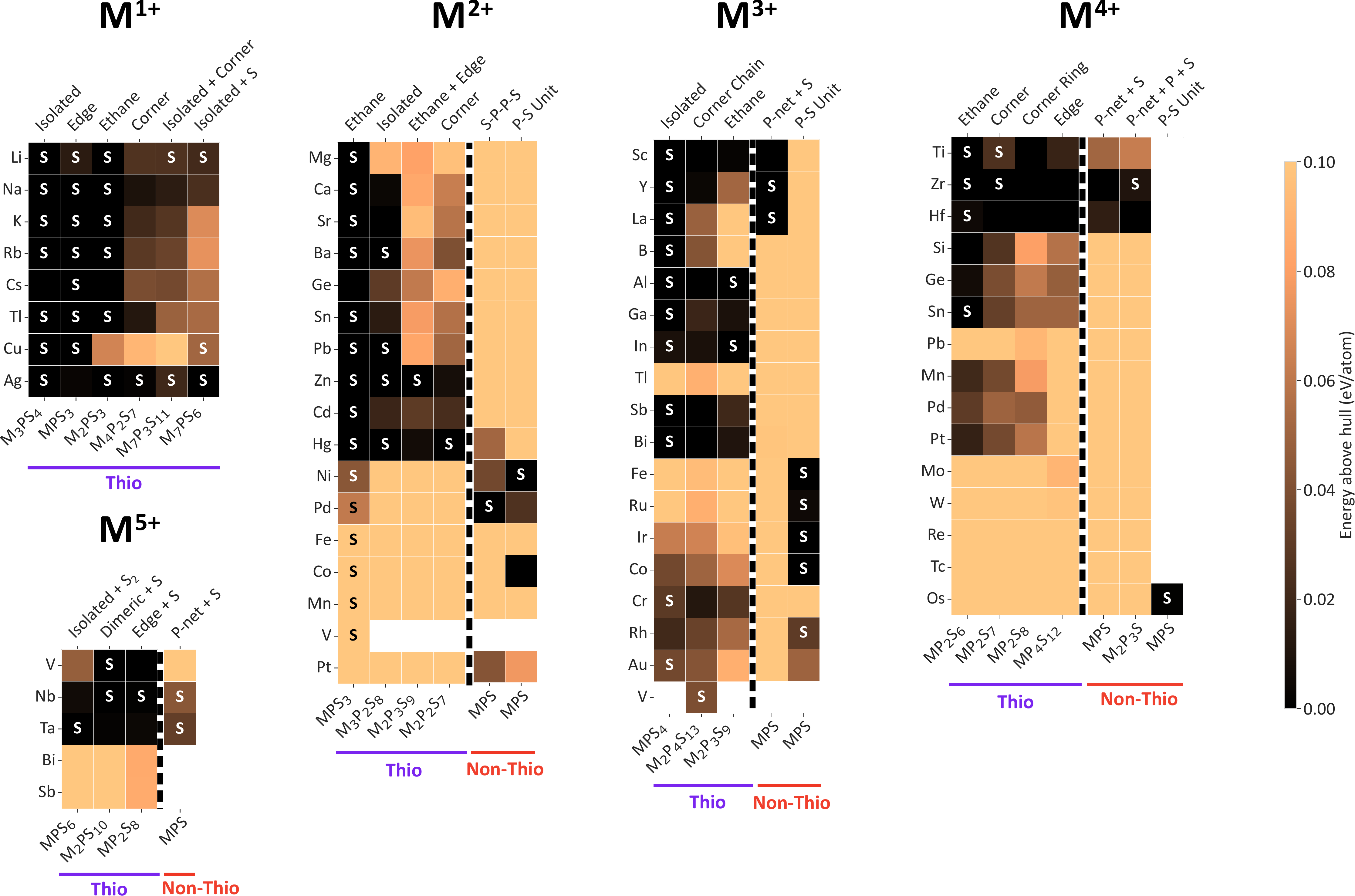}
\caption{Energy above hull heat maps of all the lowest-energy phosphosulfide polymorphs calculated in this study. Each M$^{x+}$ heat map includes phosphosulfides of metals with assumed oxidation state $x$. Along the x axes, materials with different anion motifs (top) are presented, which correspond to specific compositions (bottom). The motifs are visualized in Fig.~\ref{fig:motifs}. The color of the materials with $ E_h = \SI{100}{meV/atom}$ is also used for materials with higher energies above hull. White cells indicate that the corresponding material has not been calculated. Cells marked with an "s" indicate that the material is present in the ICSD database and is therefore identified as synthesized. The dashed vertical lines separate thiophosphate compositions ("Thio") from non-thiophosphate compositions ("Non-Thio"). The lowest-energy prototypes of each material are shown in Fig.~\ref{fig:HM_Versions}.}
\label{fig:HM_EaH}  
\end{figure*}

The only experimentally known phosphosulfide composition with $ \mathrm{P/S} > 1$ (\ce{M2P3S}, generalizable to M$_2$P$_{3-x}$S$_{1+x}$ with partial anion disorder) is found for Group 4 metals~\cite{schlechteCrystalStructureZirconium2009}. Its anion motif consists of phosphorus nets, sulfide anions, and phosphide anions. The two motifs with phosphorus nets are also the only ones where P-S bonds are absent, an otherwise ubiquitous feature in known phosphosulfides.

All known thiophosphates are charge-balanced compounds with well-defined polyanion oxidation states. Conversely, the \ce{MPS} composition includes examples of thermodynamically stable, previously synthesized compounds that are intrinsically charge imbalanced and therefore metallic, e.g., NiPS, TiPS, and NbPS. Furthermore, the \ce{MPS} composition allows for charge-neutral, semiconducting compounds from metals with various oxidation states (+2, +3, and +4).

The above discussion indicates that the P/S ratio alone does not uniquely describe the structure of the polyanions. For example, the ratio $ \mathrm{P/S} = \frac{1}{4}$ is present in three thiophosphate classes with very different motifs: Isolated \ce{[PS4]^3-} polyanions (\ce{Cu3PS4}); rings of corner-sharing \ce{PS4} tetrahedra (\ce{ZrP2S8}); and edge-sharing \ce{PS4} tetrahedra plus \ce{[S2]^2-} dimers and individual \ce{[S]^2-} anions (\ce{NbP2S8}).

Almost all known structural prototypes for ternary phosphosulfides consist of three- or two-dimensionally bonded structures~\cite{mittmannPhosphosulfideSemiconductorsOptoelectronics2024}. When considering the most stable polymorph at each composition, 76\% of phosphosulfides located on the stability hull are 3D.
For many applications where high-mobility electronic transport in all directions is desirable, 3D dimensionality is an advantage~\cite{mittmannPhosphosulfideSemiconductorsOptoelectronics2024, hoyeRoleDimensionalityOptoelectronic2022}.

\subsection{Thermodynamic stability and synthesizability}

Two factors can contribute to positive non-zero values of energy above the stability hull $E_h$ for a material \ce{M_xP_yS_z}: (i) Another polymorph with \ce{M_xP_yS_z} composition but a different structure has a lower energy; or (ii) the energy of \ce{M_xP_yS_z} lies above the plane connecting three stable compounds in the M-P-S convex hull, and therefore decomposes into these compounds. Throughout this study, we only consider the lowest-energy polymorph at each composition in our analysis, unless otherwise specified. Furthermore, we define a "stability tolerance" of $0 \leq E_h \leq \SI{100}{meV/atom}$, which we apply to exclude significantly unstable materials from most of our analysis, unless otherwise specified.

60\% of the screened phosphosulfides that lie on the stability hull have been synthesized (Fig.~\ref{fig:Synth}). The percentage of synthesized phosphosulfides rapidly drops for materials above the hull (metastable), with a synthesis rate of only 3\% when $E_\mathrm{h}$ is between 50 and $\SI{100}{meV/atom}$, and of 2\% when it is above \SI{100}{meV/atom}. 
Thus, the energy above hull metric appears to be a good descriptor for the synthesizability of phosphosulfides. Even moderate values of $E_\mathrm{hull}$ generally seem to pose a significant barrier to synthesis. In fact, sulfides and phosphides have previously been shown to be some of the material families with the lowest median $E_\mathrm{h}$ among synthesized compounds, in contrast to e.g. nitrides, where half of the metastable compounds that have been synthesized have $E_\mathrm{h} > \SI{100}{meV/atom}$~\cite{Sun2016b}.

Through our screening procedure, 19 new ternary phosphosulfide compositions with a high synthesizability likelihood ($E_\mathrm{hull} = 0$) have emerged.
They correspond to the black cells in Fig.~\ref{fig:HM_EaH} that are not marked with an "s". These unexplored materials are listed in Table \ref{tab:list_new_materials}, and include four Hf-based, three Zr-based, two Sc-based, and two Cs-based phosphosulfides. Furthermore, we find thermodynamically stable phosphosulfides (\ce{SiP2S6} and \ce{GePS3}) in the Si-P-S and Ge-P-S systems, in which no ternary compound had been known until now. 
We also find 71 more new metastable phosphosulfide compositions with moderate energies above the hull ($< \SI{50}{meV/atom}$), which may also be synthesizable but will probably present more challenges. The remaining new compositions (241) lie more than \SI{50}{meV/atom} above the stability hull, and most of them will probably be very challenging to synthesize.
Interestingly, the synthesized phosphosulfides beyond the \SI{100}{meV/atom} metastability threshold all belong to the same class of \ce{MPS3} thiophosphates based on the ethane-like polyanion coupled with open shell transition metals Fe, Co, Mn, and V~\cite{ouvrardStructuralDeterminationMPS31985}.

Heavy p-block metals often retain their two s-shell valence electrons (inert pair effect) are therefore partially oxidized in such cases. The preferred oxidation state of these metals in phosphosulfide materials can be inferred from Fig.~\ref{fig:HM_EaH} by inspecting energies above hull in their two possible oxidation states. The result is that Sn and Ge can appear both in the +2 and +4 oxidation states, whereas \ce{Tl^+}, \ce{Pb^2+}, \ce{Sb^3+}, and \ce{Bi^3+} are highly favored over \ce{Tl^3+}, \ce{Pb^4+}, \ce{Sb^5+}, and \ce{Bi^5+}.

Some compositions result in stable, synthesized materials with almost all metals with a compatible oxidation state. Such compositions are \ce{M3PS4}, \ce{MPS3}, and \ce{M2PS3} with monovalent metals, \ce{MPS3} with divalent metals, and \ce{MPS4} with trivalent p-block and d$_0$ metals (Fig.~\ref{fig:HM_EaH}).
The known non-thiophosphate structural prototypes seem to only be compatible with transition metals. Structures with P-S and S-P-P-S motifs (MPS composition) are only adopted by open-shell transition metal cations. Structures with phosphorus nets (\ce{M2P3S} and MPS compositions) are only adopted by d$_0$ cations. On the other hand, thiophosphates of open-shell transition metals based on the known structural prototypes are never on the stability hull (Fig.~\ref{fig:HM_EaH}).

\begin{figure*}[ht!]
\centering%
\includegraphics[width=0.7\textwidth]{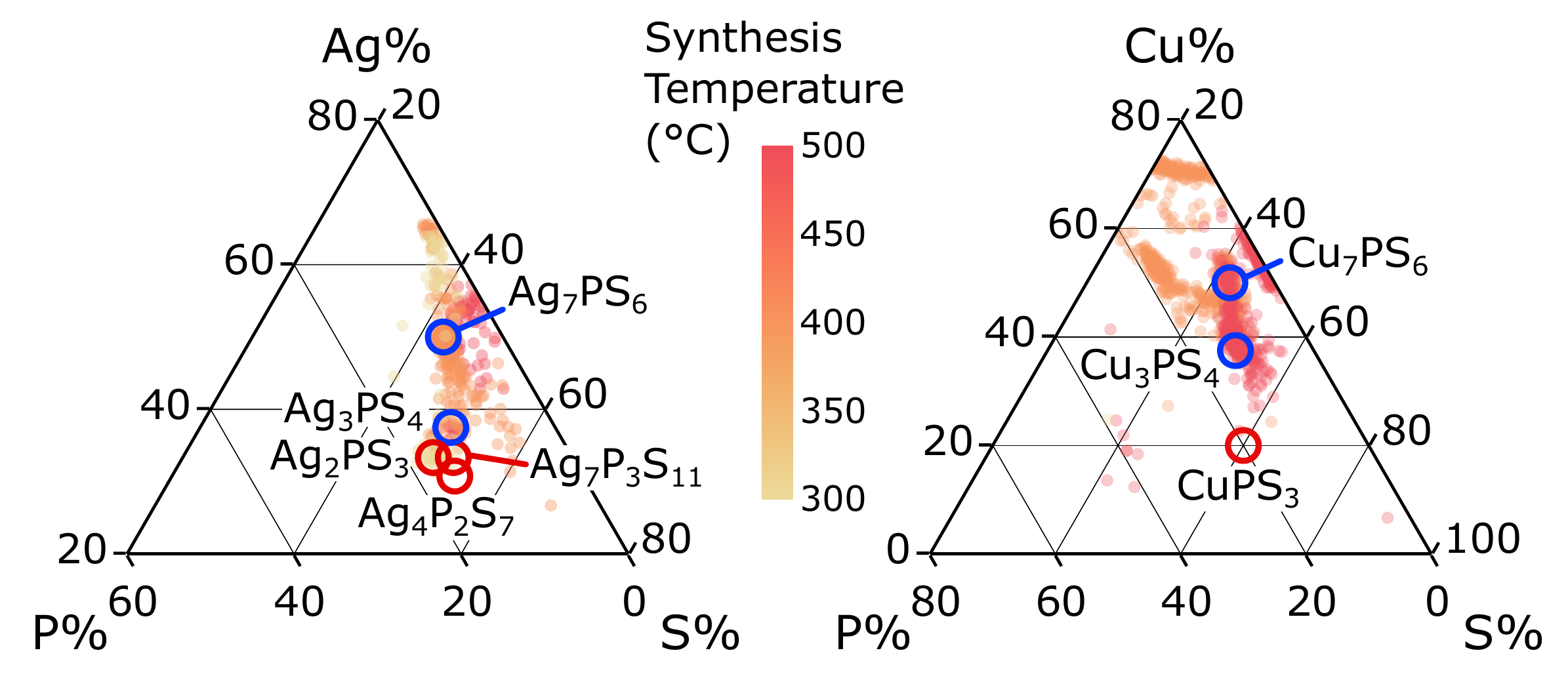}
\caption{Ag-P-S and Cu-P-S ternary systems explored by high-throughput experiments. The data points correspond to the compositions obtained from combinatorial synthesis. The three axes indicate the atomic composition of each of the three elements, normalized to a sum of 100\%.
Depending on synthesis conditions such as temperature, these data points may be single-phase compounds (e.g., \ce{Ag3PS4}) like the computationally screened ones, but also multi-phase systems (e.g., a mix of \ce{Ag2S} and \ce{Ag7PS6}) or solid solutions  (e.g. \ce{Ag_{2+3x}(P_xS)}). Single-phase compounds synthesized in this study are marked as blue circles. Other compounds calculated in this study are marked as red circles.}
\label{fig:ternary_systems}
\end{figure*}

\begin{figure}[h!]
\centering%
\includegraphics[width=1\columnwidth]{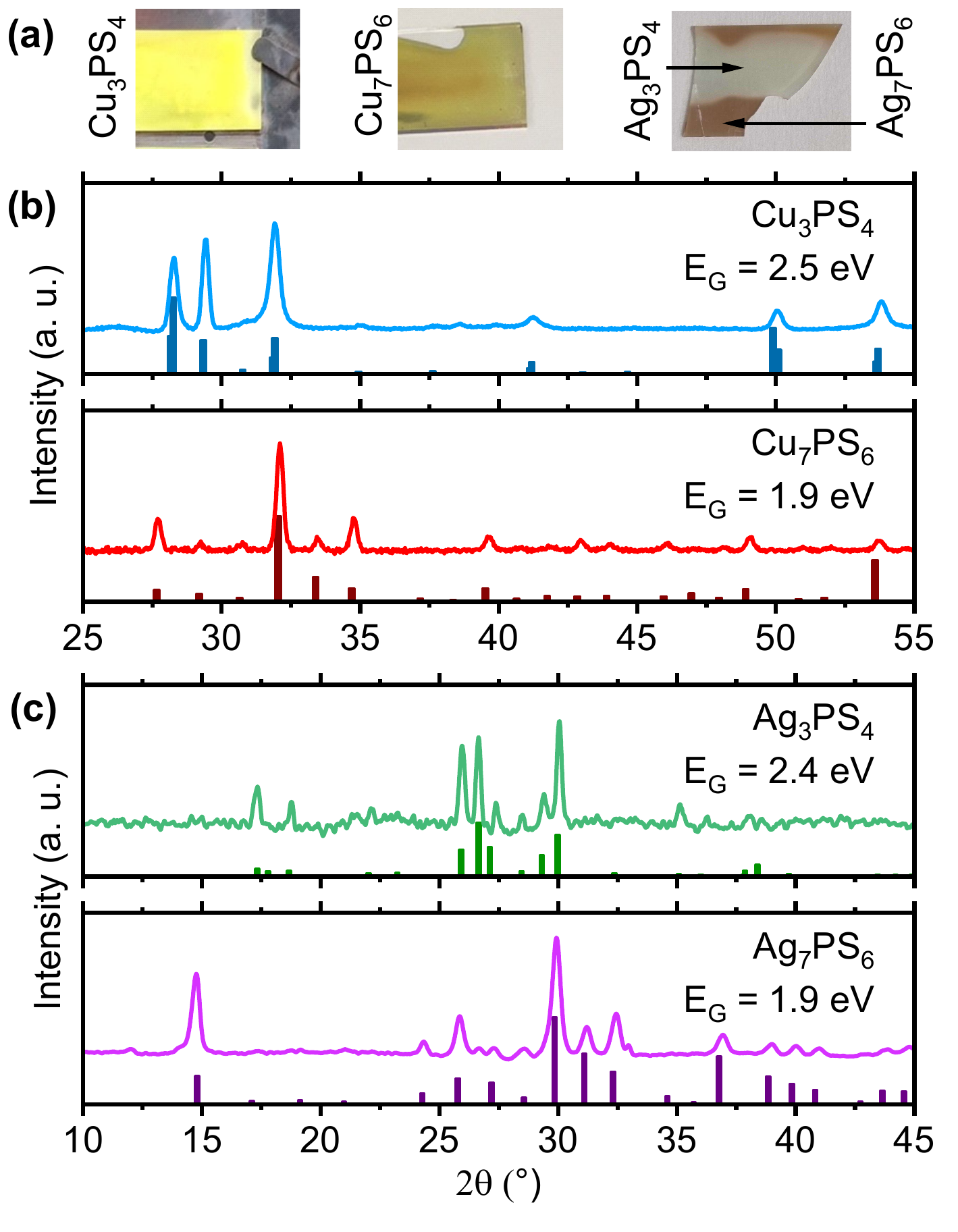}
\caption{Demonstration of thin-film synthesis of four phosphosulfides calculated in this work. (a): Photographs of the films grown on glass substrates. Note that the two Ag-P-S compounds were synthesized on the same substrate in the same combinatorial synthesis run. (b,c): Experimental XRD patterns of the four synthesized materials (continuous lines). The expected positions of the XRD reflections for the lowest-energy polymorphs calculated in this study are shown by vertical bars. Their relative heights correspond to the predicted peak intensities for a randomly oriented material, which is often not the case in thin films~\cite{Willis2022}. The experimental band gaps obtained in this work are indicated.}
\label{fig:synthesis}
\end{figure}

\begin{figure*}[ht!]
\centering%
\includegraphics[width=\textwidth]{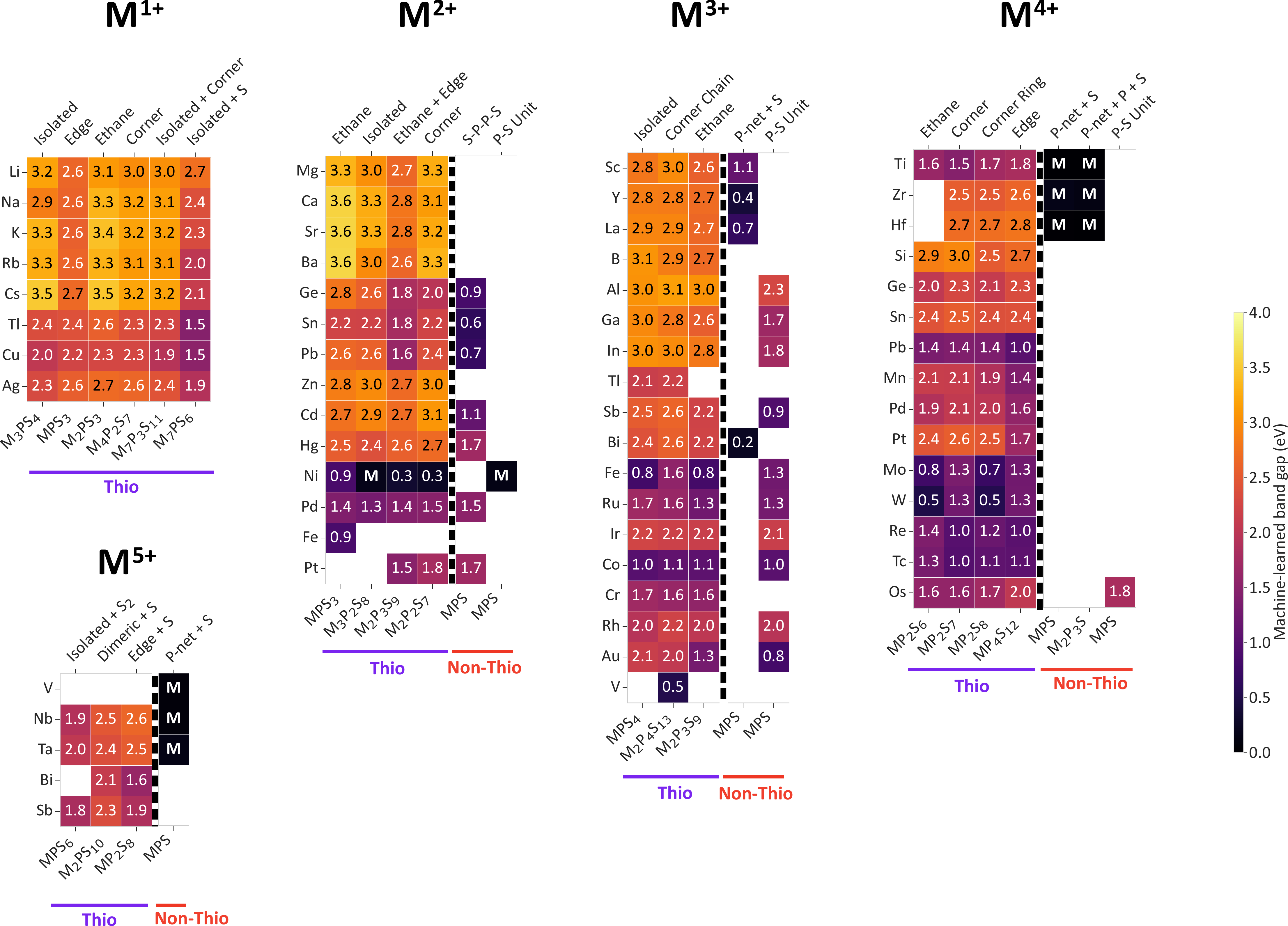}
\caption{Band gap heat maps of lowest-energy phosphosulfide polymorphs calculated in this study. Both the color and the number in each cell indicate the band gap predicted by our multi-fidelity ML model taking elemental composition and PBEsol band gap as inputs, with a mean absolute error of \SI{0.14}{eV} with respect to experimental band gaps in the full test set used when building the model (Table \ref{tab:trsl}). When a cell is marked as "M", it means that the model predicts that the compound is a metal or has a ML band gap below \SI{0.2}{eV}.
White cells indicate that a band structure for the corresponding material was not calculated.
The visualization concept is otherwise analogous to Fig.~\ref{fig:HM_EaH}}
\label{fig:HM_BG}  
\end{figure*}

\subsection{Thin-film synthesis of \ce{Cu3PS4}, \ce{Ag3PS4}, \ce{Cu7PS6}, and \ce{Ag7PS6}}

The ternary phosphosulfides reported in the ICSD database and marked as "synthesized" in Fig.~\ref{fig:HM_EaH} have been grown as bulk crystals and powders, with some examples of nanoparticle synthesis,~\cite{Yin2019} and very thin, small flakes of phosphosulfides with 2D dimensionality~\cite{yangUniversalStrategySynthesis2024}. Many applications in optics, electronics, and energy rely on large-area growth and crystallization of materials as thin films in the 0.1--1~\SI{}{\micro \meter} thickness range. As previously discussed,~\cite{mittmannPhosphosulfideSemiconductorsOptoelectronics2024} thin films of any phosphosulfide compound are still virtually unreported. In Fig.~\ref{fig:ternary_systems}, we demonstrate rapid experimental exploration of two selected ternary systems (Ag-P-S and Cu-P-S) in thin-film form using the recently proposed vacuum-based DADMARS growth technique. This high-throughput combinatorial synthesis method, described in detail elsewhere~\cite{mittmannLargeareaThinfilmSynthesis2025a}, was developed to accelerate materials discovery in experimentally challenging material systems. Each combinatorial synthesis run gives us access to about 100 unique compositions, for which the fraction of each element, crystal structure, and band gaps can be rapidly characterized using automated mapping stages.

\begin{table}[b]
\def\arraystretch{1.5}
\begin{tabular}{llllll}
\hline
\multicolumn{1}{c}{\textbf{Material}} & \multicolumn{5}{c}{\textbf{Band gap (eV)}} \\
\cline{2-6}
 & Experiment & PBEsol \hspace*{0.5mm} &  HSE \hspace*{2mm} & Fit to \hspace*{2mm} & Machine \\
  & (thin film) &  & & HSE & learned \\
\hline
\ce{Cu3PS4}      & 2.5          & 1.08     & 2.53           & 1.96                        & 2.04            \\
\ce{Cu7PS6}       & 1.9          & 0.70     &   1.68      & 1.41                        & 1.52            \\
\ce{Ag3PS4}        & 2.4*          & 1.04     & 2.23         & 1.91                        & 2.30            \\
\ce{Ag7PS6}       & 1.9          & 0.94    & 2.05          & 1.77                        & 1.91
      \\
\cline{3-6}
\textbf{MAE}      &              & \textbf{1.24}   & \textbf{0.14}   & \textbf{0.41}               & \textbf{0.24}
      \\
\hline
\end{tabular}
\caption{Band gaps of the four synthesized compounds using different methods. The experimental band gap of \ce{Ag3PS4} is marked with an asterisk because the optically determined band gap is \SI{2.7}{eV}, but optical transitions in \ce{Ag3PS4} were found to be dipole-forbidden up to \SI{0.3}{eV} above its fundamental band gap~\cite{Han2021}. HSE band structures are shown in Fig.~\ref{fig:HSE_Bands}. The mean absolute error (MAE) of each method with respect to our in-house experimental band gaps is indicated in the bottom row.}
\label{tab:bandgaps}
\end{table}

We were able to synthesize four single-phase thin-film compounds (\ce{Cu3PS4}, \ce{Cu7PS6}, \ce{Ag3PS4} and \ce{Ag7PS6}) with only four combinatorial experiments and without pre-existing knowledge of synthesis recipes. Their visual appearance and XRD patterns are shown in Fig.~\ref{fig:synthesis}. Remarkably, two single-phase compounds (\ce{Ag3PS4} and \ce{Ag7PS6}) were obtained in a single combinatorial synthesis run. \ce{Cu3PS4} and \ce{Ag3PS4} feature the isolated \ce{PS4} tetrahedron motif, whereas \ce{Cu7PS6} and \ce{Ag7PS6} feature isolated \ce{PS4} tetrahedra combined with \ce{S^2-} anions. These four materials are predicted to be on the stability hull, except for \ce{Cu7PS6}, which lies \SI{51}{meV/atom} above (Fig.~\ref{fig:HM_EaH}) but may be disorder-stabilized similar to \ce{Cu7PSe6}~\cite{Foster2013}.

The synthesized films are air-stable semiconductors with band gaps in the visible. Visually darker films correspond to lower band gaps (Fig.~\ref{fig:synthesis}(a)). Standard band gap extraction methods based on optical transmission and reflection measurements (Fig.~\ref{fig:abs}) yield the experimental band gaps shown in Table~\ref{tab:bandgaps}, with an estimated uncertainty of \SI{\pm 0.1}{eV} (sources of error are discussed in the SI). For comparison, we also show in-house calculated band gaps by four methods, two based on DFT calculations (PBEsol and HSE functionals), and two based on data-driven corrections to the PBEsol gaps. The two correction methods are: (i) Polynomial fitting between PBEsol band gaps and HSE band gaps; (ii) Machine learning (ML) of band gaps from PBEsol fidelity to experimental fidelity. As expected, the PBEsol level suitable for high-throughput calculations highly underestimates the band gaps of these compounds, whereas the more computationally demanding HSE band gaps are in very good agreement with experiment (Table~\ref{tab:bandgaps}). Among the two methods to improve the quality of PBEsol band gaps, the multifidelity ML approach performs better than the simple PBEsol-to-HSE fitting approach, with a mean absolute prediction error (MAE) of \SI{0.24}{eV} with respect to experiment. This error is expected to improve with a larger statistical sample of materials, since the MAE of the same model on a much larger test dataset was \SI{0.17}{eV} (see Methods section). Thus, rapid calculation of PBEsol band gaps followed by their ML-based extrapolation to experimental fidelity can be a useful for materials screening, where higher-accuracy first-principles methods would be too computationally demanding.

\subsection{Band gaps}
With a more detailed awareness of the quality of multifidelity ML band gap model, we translate the PBEsol band gaps of our calculated phosphosulfides to the experimental fidelity level (Fig.~\ref{fig:HM_BG}).
Band gaps span a wide range from metallic to around 4~eV. Overall, the widest band gaps are found in alkali and alkaline earth metals, while open-shell transition metals often have narrow band gaps. Moving horizontally in each of the heat maps in Fig.~\ref{fig:HM_BG}, there are some band gap changes across compositions and polyanion motifs at fixed metal cation. However, the three most common motifs (isolated \ce{PS4}, corner-sharing \ce{PS4}, and ethane) tend to give similar band gaps for the same M.

\begin{figure}[ht!]
\centering%
\includegraphics[width=\columnwidth]{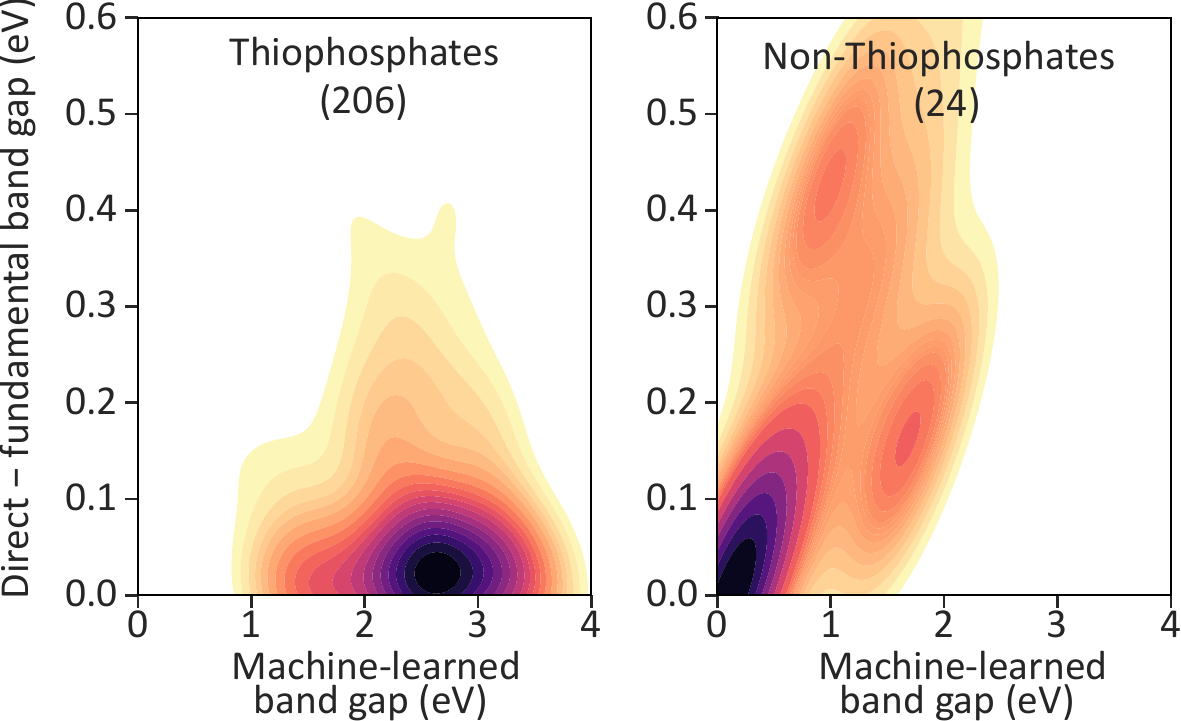}
\caption{Contour plots showing the (fundamental) band gap predicted by our multi-fidelity ML model versus the difference between the lowest direct band gap and the fundamental gap, as derived from PBEsol band structures. This difference is greater than zero if the fundamental gap is indirect.
Darker colors indicate a higher density of materials. The plot on the left only includes thiophosphates. The plot on the right only includes non-thiophosphates, as defined in Fig.~\ref{fig:motifs}. Only lowest-energy polymorphs within the stability tolerance are included. The number of material entries used in these plots is shown in parentheses.}
\label{fig:BG_vs_indirectedness}
\end{figure}

There are important qualitative differences between thiophosphates and non-thiophosphates.
Thiophosphates occur most frequently in the 2.3--3.0~eV band gap range (Fig.~\ref{fig:BG_vs_indirectedness}) with a preference for slightly indirect gaps. Interestingly, the maximum in our ML-predicted band gap distribution matches the corresponding maximum determined for all known sulfide compounds~\cite{Varley2017}, giving further credibility to our proposed method for accelerated band gap prediction.
The median difference between the lowest direct gap and the fundamental gap in thiophosphates is \SI{23}{meV} when only considering indirect gap materials, and \SI{14}{meV} when including both direct and indirect gap materials. Thus, even in indirect gap thiophosphates, the direct gap is typically within thermal energy of the fundamental gap at room temperature. While indirect gaps have historically been considered a drawback for optoelectronic applications, an indirect gap just below a direct one is generally benign or even beneficial for optoelectronic energy conversion when the device is limited by charge collection~\cite{kirchartzDecreasingRadiativeRecombination2017,kangsabanikIndirectBandGap2022}. Thiophosphates are therefore well positioned as potential optoelectronic materials for a variety of applications where band gaps in the yellow-blue region are desirable, such as light emitting diodes, photocatalysis, and solar fuels. According to Fig.~\ref{fig:HM_BG}, thiophosphates seem to have a very low propensity for metallic band structures.

On the other hand, non-thiophosphates have a more complex band gap distribution with three local maxima (Fig.~\ref{fig:BG_vs_indirectedness}) that are characteristic of the three anion motifs identified in Fig.~\ref{fig:motifs}.
The first local maximum is characteristic of materials with the P-S unit (see Fig.~\ref{fig:nonthio_BG}) and is located at (indirect) band gaps around \SI{1.2}{eV}, where the direct band gap lies around 0.4~eV above the indirect one. This maximum remains very clearly defined when the number of materials is increased by a factor 4 by also including higher-energy polymorphs based on the P-S unit (Fig.~\ref{fig:nonthio_BG}).

The second local maximum is found when approaching the metallic limit (\SI{0}{eV}). This maximum is characteristic of non-thiophosphates based on P nets and \ce{S^{2-}} anions, either with or without separated \ce{P^{3-}} anions (Fig.~\ref{fig:nonthio_BG}). Here, we should note that the ML model often predicts the opening of very small band gaps ($< \SI{0.2}{eV}$) in phosphosulfides that are metallic at the PBEsol level. This is also observed in some of the MPS compositions that we predicted to be charge-imbalanced in Table~\ref{tab:bandgaps} and therefore intrinsically metallic, indicating an unphysical ML prediction. Since the uncertainty in the ML-predicted band gap is around \SI{0.2}{eV} (Tables~\ref{tab:bandgaps}, \ref{tab:trsl}), we label all materials with predicted band gaps below this threshold as metallic in Fig.~\ref{fig:HM_BG}. 
Distinguishing between true metals and narrow-gap semiconductors is likely one of the most challenging tasks for the multi-fidelity ML model and, more generally, for any method trying to determine the band gap of a material where a semilocal DFT functional yields a metallic band structure. Although the ML model is trained on experimental band gaps, experimental methods themselves often struggle to distinguish between metals, semimetals, and semiconductors with very narrow gaps~\cite{crovettoCu3xPSemiconductorMetal2023}.

The third local maximum for non-thiophosphate is centered around \SI{1.7}{eV} band gap with moderately indirect gaps (on average \SI{0.15}{eV} below the lowest direct gap). This maximum arises from the materials with the S-P-P-S motif. However, this motif only has one known prototype and three lowest-energy polymorphs within the stability threshold. When higher-energy materials are included (Fig.~\ref{fig:nonthio_BG}) new maxima appear, indicating that this third maximum is not an intrinsic feature of the S-P-P-S motif.

The combined band gap distributions of thiophosphates and non-thiophosphates reveal that ternary metal phosphosulfides as a whole cover almost the full spectrum of band gap combinations, from metallic structures to low, moderate, and wide band gaps with various degrees of indirectedness. Only ultra-wide band gaps above \SI{3.6}{eV} are not represented. Moving to quaternary compounds is expected to further expand possiblities for materials design. The space of non-thiophosphate compounds features particularly diverse bonding patterns and electronic properties in spite of its small size. Thus, this family appears particularly promising for further discovery of materials with unexpected properties. Many chemically plausible non-thiophosphates are predicted to be thermodynamically unstable by our calculations (Fig.~\ref{fig:HM_EaH}). However, these calculations rely on a small range of structural prototypes which reflect the very limited research attention they have received so far, with no experimental reports of any quaternary non-thiophosphates~\cite{mittmannPhosphosulfideSemiconductorsOptoelectronics2024}, nor of non-thiophosphates with metals outside the d block (Fig.~\ref{fig:HM_EaH}). It remains to be seen whether many synthesizable non-thiophosphates are still awaiting discovery.

\section{Conclusions}
We evaluated the thermodynamic stability and synthesizability of nearly a thousand ternary phosphosulfides by high-throughput DFT, discovering 19 previously unknown materials that lie on the stability hull.
We found that the synthesizability of phosphosulfides is highly correlated with their energy above hull. Although there are a few examples of synthesized phosphosulfides lying higher than \SI{100}{meV/atom} above the stability hull, synthesis of highly metastable phosphosulfides appears to be more difficult than in the case of oxides and especially nitrides.

We developed a multi-fidelity machine learning model to rapidly predict experimentally calibrated band gaps with elemental composition and PBEsol band gaps as the sole inputs.
By including our in-house experimental samples in the validation of the multi-fidelity model, we conclude that the model accuracy is not far from the typical experimental error in measuring the band gap of a synthesized sample without any prior knowledge.

The amphoteric nature of phosphorus, acting either as a cation or as an anion depending on chemical context, results in two very different classes of phosphosulfides: Thiophosphates and non-thiophosphates.
The transition from one class to another occurs at a P/S ratio of 1, coinciding with a switch in the phosphorus oxidation state from positive to negative.
Structural motifs, compatible metal cations, band gap distributions, tendency to indirect band gaps, and tendency to metallicity are completely different in thiophosphates and non-thiophosphates.
We expect that these insights can be used to intelligently search for novel phosphosulfides with entirely new compositions, new structures, and higher complexity (quaternary compounds and beyond) in the future.

Importantly, the high throughput of our workflow is not limited to computational aspects but it extends to the experimental domain, including both synthesis and characterization. This feature was enabled by a unique thin-film synthesis technique (DADMARS) that was specifically designed for accelerated, yet high-quality synthesis of experimentally challenging material classes~\cite{mittmannLargeareaThinfilmSynthesis2025a}. In practice, the high-throughput experimental setup allowed us to synthesize four phosphosulfides that had not previously been grown as thin films after only four combinatorial synthesis runs. This work demonstrates that accelerated materials development workflows combining theory, AI, and experiment are within reach even for "difficult" chemical systems, as long as a suitable experimental setup has been designed.

\section*{Methods}
\subsection{First-principles calculations}
First-principles calculations were performed using DFT with the projector-augmented wave method~\cite{Blochl1994} as implemented in the Vienna Ab-Initio Simulation Package (VASP)~\cite{kresseInitioMolecularDynamics1993,kresseInitioMoleculardynamicsSimulation1994,kresseEfficiencyAbinitioTotal1996,kresseEfficientIterativeSchemes1996,kresseUltrasoftPseudopotentialsProjector1999}. We only considered ternary metal phosphosulfides in structures that had been experimentally reported for at least one phosphosulfide material, according to the Inorganic Crystal Structure Database (ICSD)~\cite{zagoracRecentDevelopmentsInorganic2019}. In a first screening step, we decorated each known structural prototype with metals with a compatible oxidation state and calculated the relaxed geometry and total energy of each unique compound. The initial geometries were taken from the Material Project database.~\cite{Jain2013}
A full list of the prototypes used in this work, criteria for their selection, and their corresponding Materials Project ID, are available in the SI (Table~\ref{tab:mp-id}). All calculations were submitted using the SLURM frontend MyQueue~\cite{mortensenMyQueueTaskWorkflow2020}.

Based on the preliminary tests shown in the SI, the PBESol functional was selected for all calculations reported in this paper, unless otherwise specified~\cite{perdewRestoringDensityGradientExpansion2008}. The structure relaxation calculations were run on $\Gamma$-centered uniform \textit{k}-point grids with a 680~eV plane wave energy cutoff and 5 \textit{k}-points per \AA$^{-1}$ along each reciprocal lattice direction (VASP kspacing parameter: 0.22). The lattice parameters and atomic positions were relaxed with convergence criteria of less than \SI{10}{meV/\AA} forces on all atoms, and less than $10^{-6}\,\mathrm{eV}$ total energy difference between steps. All calculations were spin polarized.

The energies above the stability hull were determined by constructing convex hulls for the 50 ternary M-P-S systems considered in this study using pymatgen, as detailed in the SI. The structural dimensionality of all the calculated materials was determined using the algorithm by Larsen et al.~\cite{larsenDefinitionScoringParameter2019} as implemented in pymatgen~\cite{Ong2013}.

For the materials found to lie within \SI{200}{meV/atom} of the stability hull, we proceeded to a non self-consistent calculation of their electronic band structure. These calculations were run along high symmetry paths in the Brillouin zone, as defined in Ref.~\cite{setyawanHighthroughputElectronicBand2010}, with 20 \textit{k}-points for each segment in the path and 680~eV plane wave energy cutoff. From the calculated band structures, we extracted the fundamental band gap and lowest direct band gap of each material using pymatgen. For spin-polarized band structures (about 9\% of all band structures within the stability tolerance) we kept the spin channels separate (no spin flipping) and quoted the fundamental and lowest direct band gaps from the spin channel with the lowest fundamental band gaps. For the four materials that we synthesized as thin films, we performed complementary relaxation and band structure calculations using the hybrid HSE06 functional~\cite{Heyd2003,Krukau2006} with a plane wave energy cutoff of 500~eV or higher, and 20 \textit{k}-points for each segment in the path. Further details about the first-principles DFT calculations are given in the SI.

\subsection{Machine/statistical learning}

We employed two statistical methods to correct PBEsol band gaps.
The first method is a simple second-order polynomial regression between HSE and PBEsol band gaps, using a dataset of 39 ternary phosphosulfides for which HSE band gaps were available on the SNUMAT database~\cite{Kim2020b} and the PBEsol band gaps were calculated in this work. This approach is analogous to the HSE to PBE regression method employed in previous work~\cite{mittmannPhosphosulfideSemiconductorsOptoelectronics2024}. The best-fit parameters and fit quality are shown in Fig.~\ref{fig:HSE Fit}.

The second method is a multi-fidelity deep learning approach based on permutation-invariant transformers. We developed this method to predict the experimental band gaps of inorganic materials from two inputs: (i) chemical composition, and (ii) a calculated band gap at one of six possible fidelity levels, corresponding to different DFT functionals (PBE, PBE+U, PBEsol, SCAN, GLLB-SC, and HSE). For this work, the input fidelity level of interest is PBEsol. The model was partially inspired by Ref.~\cite{chenLearningPropertiesOrdered2021}, with the important difference that here structural information is not needed.

Two distinct (sub)models were developed to obtain a prediction-ready multi-fidelity platform. The base model uses chemical composition as the only input to predict the band gap at a given fidelity level (e.g, PBEsol).
The translation model is a two-input model that converts the band gap between fidelity levels. It uses chemical composition and a known band gap at a given fidelity level (e.g., PBEsol) to predict the band gap at another level (e.g., experimental).

The PBE and PBE+U datasets (91,920 and 21,917 entries respectively) were obtained from the Materials Project database~\cite{Jain2013}. The PBEsol and SCAN datasets (113,106 and 112,722 entries) were obtained from the database also used to construct our convex hull diagrams~\cite{schmidtDataset175kStable2022}. The GLLB-SC dataset (2,395 entries) was generated in~\cite{Castelli2015} and obtained through MPContribs~\cite{huckUserApplicationsDriven2016}. The HSE dataset (10,482 entries) was obtained from the SNUMAT database.~\cite{Kim2020b} Finally, the experimental dataset (4,605 entries) was compiled by the authors of a previous ML model~\cite{chenLearningPropertiesOrdered2021} based on an earlier compilation~\cite{Zhuo2018} and released with the accompanying paper. About half of the materials in this dataset are metals, while the other half have non-zero band gaps.
When preparing the datasets for training, we retained only the band gap of the lowest-energy polymorph at each composition, as necessary for models not incorporating structural information~\cite{Bartel2020}. Details are available in the SI.

The mean absolute error (MAE) for the base model to predict experimental band gaps from composition only is \SI{0.37}{eV} (Table~\ref{tab:base}).
The 5-fold cross-validation MAE for the translation model to predict an experimental band gap from composition and a PBEsol band gap is \SI{0.17}{eV} (Table~\ref{tab:trsl}). 

\subsection{Experiments}

In line with the rest of this study, the synthesis, experimental characterization, and experimental data management were conducted with high-throughput methods.
Thin-film combinatorial libraries~\cite{gregoireCombinatorialSynthesisAIdriven2023} were synthesized by directional-and-diffuse multi-anion reactive sputtering (DADMARS), a technique presented in depth in a previous publication~\cite{mittmannLargeareaThinfilmSynthesis2025a}. Briefly, a metallic (M) sputter source (Cu or Ag), reactive \ce{PH3} gas, and a cracked sulfur beam were combined in an Ar discharge to deposit M-P-S films with intentional gradients in elemental composition across a large growth area. The synthesis pressure was \SI{0.67}{Pa}, the synthesis temperature was varied between \SI{300}{\celsius} and \SI{500}{\celsius}, and the fluxes of M, P, and S resulting from these sources were varied from one synthesis run to another to sample different portions of the Cu-P-S and Ag-P-S systems. These fluxes were intentionally inhomogeneous in space to obtain the desired composition gradients allowing for rapid combinatorial exploration. Elemental composition was measured by energy-dispersive x-ray spectroscopy (EDX). Structural information was inferred from x-ray diffraction (XRD). Band gaps were determined from optical transmission and reflection spectra. Further details of the experimental characterization are given in the SI. High-throughput experimental data was managed and analyzed with customized workflows
in a NOMAD Oasis database~\cite{scheidgenNOMADDistributedWebbased2023} with human intervention mainly limited to validation steps.

\begin{acknowledgments}
We thank Benjamin H. Sjølin for help with setting up the initial workflow and Finja Tadge for running some of the DFT calculations. We acknowledge Mikkel N. Schmidt for useful discussions on the machine learning model. This work was co-funded by the European Union (ERC, IDOL, 101040153). Views and opinions expressed are however those of the authors only and do not necessarily reflect those of the European Union or the European Research Council. Neither the European Union nor the granting authority can be held responsible for them. This work was supported in part by a research grant (42140) from VILLUM FONDEN. We also acknowledge support from the Novo Nordisk Foundation Data Science Research Infrastructure 2022 Grant: A high-performance computing infrastructure for data-driven research on sustainable energy materials, Grant no. NNF22OC0078009.
\end{acknowledgments}

\section*{Data availability}
The computational and experimental data generated from this study, including all DFT calculations and a database of computed properties, is available in the NOMAD database at \url{https://dx.doi.org/10.17172/NOMAD/2026.01.22-2}. The trained multi-fidelity machine learning model used to translate band gaps is available on  \url{https://github.com/nkrygernelson/FETT}.

\bibliography{PSdiscovery}

\newpage
\clearpage


\onecolumngrid
\begin{center}
\begin{large}
\textbf{SUPPORTING INFORMATION \\}
\vspace{0.5cm}
\textbf{AI-enhanced discovery and accelerated synthesis of metal phosphosulfides \\}
\end{large}
\vspace{0.5cm}
Javier Sanz Rodrigo, Nicholas A. Kryger-Nelson, Lena A. Mittmann, \\ Eugène Bertin, Ivano E. Castelli, Andrea Crovetto*
\end{center}

\vspace{1cm}

\setcounter{page}{1}
\renewcommand*{\thepage}{S\arabic{page}}
\renewcommand{\thefigure}{S\arabic{figure}}
\renewcommand{\thetable}{S\arabic{table}}

\setcounter{figure}{0}
\setcounter{section}{0}


\section{Extended computational details}

\subsection{First-principles calculations}

To select an appropriate exchange-correlation functional for our high-throughput study, we structurally relaxed 14 experimentally reported phosphosulfides using the PBE~\cite{Perdew1996}, PBEsol~\cite{perdewRestoringDensityGradientExpansion2008} and R$^2$SCAN~\cite{furnessAccurateNumericallyEfficient2020} functionals. Table \ref{tab:lattice_parameters} lists the error of the calculated lattice parameters with respect to the experimental ones, taken from the ICSD. PBEsol yields the lowest error of ($-0.3 \pm 1.0$)\% with respect to experiment (Table~\ref{tab:lattice_parameters_2}) and was therefore selected for this study.

Initial magnetic moments were set to zero except for the case of open-shell transition metal phosphosulfides, where we performed separate calculations with the magnetic moment of the metal initialized to zero and to its high-spin state. We then selected the calculation results with the lowest total energy for further processing. We did not attempt to incorporate Hubbard $U$ corrections~\cite{Anisimov1991} to transition metals for two reasons. First, physically grounded values of $U$ for the different metals are not known \textit{a priori}, especially for understudied material systems like phosphosulfides. Second, the ML model we developed to correct PBEsol band gaps (described later) was trained by comparing experimental band gaps to a PBEsol band gap database~\cite{schmidtDataset175kStable2022} where Hubbard corrections were not used for any compound. A similar argument can be used to justify our choice of neglecting relativistic effects such as spin-orbit coupling, which were not used in the PBEsol band gap training database~\cite{schmidtDataset175kStable2022}.

\begin{table}[h!]
\centering
\begin{tabular}{llllllllll}
\hline
\multicolumn{1}{c}{} & \multicolumn{3}{c}{\textbf{PBE vs.}}     & \multicolumn{3}{c}{\textbf{PBEsol vs.}}                                   & \multicolumn{3}{c}{\textbf{R$^2$SCAN vs.}}  \\
\multicolumn{1}{c}{} & \multicolumn{3}{c}{\textbf{experiment}}     & \multicolumn{3}{c}{\textbf{experiment}}                                   & \multicolumn{3}{c}{\textbf{experiment}}  \\
\textbf{Material}  & \textbf{a} & \textbf{b} & \textbf{c} & \multicolumn{1}{l}{\textbf{a}} & \multicolumn{1}{l}{\textbf{b}} & \multicolumn{1}{l}{\textbf{c}} & \multicolumn{1}{l}{\textbf{a}} & \multicolumn{1}{l}{\textbf{b}} & \multicolumn{1}{l}{\textbf{c}} \\ \hline
\renewcommand{\arraystretch}{1.5}
\ce{Li7P3S11}             & 2.01\%     & 4.38\%     & -0.27\%    & 1.56\%                & 0.99\%                & -1.92\%               & 3.83\%                & 1.52\%                & -1.68\%               \\
\ce{Th(PS3)2}             & 1.28\%     & 1.28\%     & 0.75\%     & 0.19\%                & 0.19\%                & -1.18\%               & 1.68\%                & 1.68\%                & 0.34\%                \\
\ce{P2Pb3S8}              & 1.47\%     & 1.47\%     & 1.47\%     & -0.90\%               & -0.90\%               & -0.90\%               & 1.08\%                & 1.08\%                & 1.08\%                \\
\ce{Ba3(PS4)2}            & 0.53\%     & -0.13\%    & 1.09\%     & -1.62\%               & -2.19\%               & -0.70\%               & 0.08\%                & -0.50\%               & 0.60\%                \\
\ce{SbPS4}                & 1.82\%     & 6.09\%     & 5.85\%     & 1.12\%                & -0.10\%               & -0.59\%               & 0.79\%                & 3.07\%                & 2.88\%                \\
\ce{InPS4}                & 2.72\%     & 2.72\%     & 2.43\%     & 1.37\%                & 1.37\%                & 0.72\%                & 2.39\%                & 2.39\%                & 1.52\%                \\
\ce{Na3PS4}               & 0.46\%     & 0.46\%     & 0.48\%     & -1.19\%               & -1.19\%               & -1.06\%               & -1.06\%               & -1.06\%               & -0.38\%               \\
\ce{Tl3PS4}               & 2.62\%     & 1.82\%     & 3.18\%     & -0.42\%               & -1.62\%               & -0.72\%               & 1.24\%                & -0.05\%               & 0.38\%                \\
\ce{Al2(PS3)3}            & 0.57\%     & 0.55\%     & 3.89\%     & -1.00\%               & -1.00\%               & 1.38\%                & -0.14\%               & -0.08\%               & 2.34\%                \\
\ce{Ag4P2S7}              & 1.12\%     & 1.14\%     & 1.05\%     & -0.52\%               & -1.50\%               & -1.02\%               & 0.85\%                & 0.09\%                & 0.79\%                \\
\ce{Hg2P2S7}              & 2.52\%     & 2.19\%     & 2.60\%     & 0.14\%                & 0.12\%                & 0.03\%                & 0.86\%                & 2.27\%                & 2.26\%                \\
\ce{CdPS3}                & 1.26\%     & 1.28\%     & 1.93\%     & -0.59\%               & -0.54\%               & 0.77\%                & 0.52\%                & 0.55\%                & 3.46\%                \\
\ce{CsPS3}                & 1.15\%     & 1.31\%     & 1.35\%     & -0.17\%               & 0.06\%                & 0.21\%                & 1.33\%                & 1.55\%                & 0.98\%                \\
\ce{HgPS3}                & 1.68\%     & 1.66\%     & 2.46\%     & -0.35\%               & -0.09\%               & 1.30\%                & 1.31\%                & 1.38\%                & 3.04\%  \\
\hline             
\end{tabular}
\caption{Error in the calculated lattice parameters \textbf{a}, \textbf{b}, and \textbf{c} of a test set of phosphosulfide materials with respect to experimental values. A positive value indicates that the calculated lattice parameter is larger than the experimental one. \label{tab:lattice_parameters}}
\end{table}

\begin{table}[h!]
\begin{tabular}{cccc}
\hline
{\color[HTML]{000000} \textbf{}} & {\color[HTML]{000000} \textbf{PBE}} & {\color[HTML]{000000} \textbf{PBESol}} & {\color[HTML]{000000} \textbf{R$^2$SCAN}} \\ \hline
{\color[HTML]{000000} \textbf{Average}}    & {\color[HTML]{000000} 1.80\%}       & {\color[HTML]{000000} -0.30\%}         & {\color[HTML]{000000} 1.10\%}          \\
{\color[HTML]{000000} \textbf{Std. Dev.}}  & {\color[HTML]{000000} 1.36\%}       & {\color[HTML]{000000} 0.98\%}          & {\color[HTML]{000000} 1.24\%}    \\
\hline      
\end{tabular}
\caption{Average error and standard deviation of the error in the lattice parameters calculated with the three functionals in Table~\ref{tab:lattice_parameters}. \label{tab:lattice_parameters_2}}
\end{table}

Most of the known ternary phosphosulfides have 3D structural dimensionality, i.e., they feature primary bonding networks along all three dimensions (Fig.~\ref{fig:motifs}). Very few phosphosulfides have structural dimensionality lower than 2D~\cite{mittmannPhosphosulfideSemiconductorsOptoelectronics2024}. Hence, one may expect that van der Waals interactions do not significantly affect the total energy of most phosphosulfides, so that neglecting these interactions may be tolerated in a high-throughput study. Test calculations on 15 phosphosulfides with 3D and 2D structures (Table~\ref{tab:vdW}) revealed that neglecting van der Waals interactions results in less negative formation energies by an average of \SI{68}{meV/atom}. However, the standard deviation of the difference between formation energies with and without van der Waals interactions is small (\SI{9}{meV/atom}). Since changes in the convex stability hull from one case to the other are mainly driven by this deviation, we chose not to model van der Waals interactions.

\begin{table}[h!]
\begin{tabular}{llllll}
\hline
\multicolumn{1}{l}{} & \textbf{Structural} & \textbf{PBESol formation } & \textbf{PBESol+vdW formation } & \textbf{Formation energy} \\
\multicolumn{1}{l}{\textbf{Material}} & \textbf{dimensionality} & \textbf{energy (eV/atom)} & \textbf{energy (eV/atom)} & \textbf{difference (eV/atom)} \\\hline
\ce{B2P3S9}                                & 2D           & -0.171         & -0.240             & -0.068       \\
\ce{B2P4S13}                               & 2D           & -0.243         & -0.304             & -0.061       \\
\ce{B2PS}                                  & 3D           & 0.616          & 0.547              & -0.068       \\
\ce{BP2S7}                                 & 2D           & -0.160         & -0.219             & -0.059       \\
\ce{BPS}\_v1                              & 3D           & -0.118         & -0.187             & -0.069       \\
\ce{BPS}\_v2                             & 3D           & 0.462          & 0.406              & -0.056       \\
\ce{BPS}\_v3                              & 3D           & 0.182          & 0.112              & -0.070       \\
\ce{BPS}\_v4                            & 2D           & -0.123         & -0.193             & -0.070       \\
\ce{BPS3}                                 & 2D           & -0.239         & -0.309             & -0.070       \\
\ce{BPS4}\_v1                            & 2D           & -0.318         & -0.395             & -0.077       \\
\ce{BPS4}\_v2                           & 2D           & -0.346         & -0.418             & -0.072       \\
\ce{BPS4}\_v3                              & 2D           & -0.230         & -0.292             & -0.062       \\
\ce{BPS4}\_v4                              & 3D           & -0.331         & -0.419             & -0.088       \\
\ce{BPS4}\_v5                             & 2D           & -0.168         & -0.243             & -0.075       \\
\ce{BPS4}\_v6                               & 3D           & -0.288         & -0.338             & -0.050 \\
\hline
\end{tabular}
\caption{Formation energies per atom of the ternaries in the B - P - S material space with and without van der Waals corrections (DFT-D3). Van der Waals corrections were also applied to the elemental references to obtain consistent formation energies. The average effect of the Van der Waals correction is a decrease in formation energy by -67.6 $\pm$ 9 meV/atom. \label{tab:vdW}}
\end{table}

To derive convex stability hulls for the 50 M-P-S systems investigated in this work, we used the calculated formation energies of ternary phosphosulfides and of M-P, M-S, and P-S binaries. The energies of ternary phosphosulfides were calculated in this work. The energies of the elements and of most binaries were obtained from an existing computational database based on the PBEsol functional~\cite{schmidtDataset175kStable2022}. For the binaries that were absent from the PBEsol database but that were close to the convex hull on the (PBE-based) Materials Project, we calculated the PBEsol total energy using the same calculation parameters as for ternary phosphosulfides and included them in our convex hulls.

\begin{figure*}[ht!]
\centering%
\includegraphics[width=\textwidth]{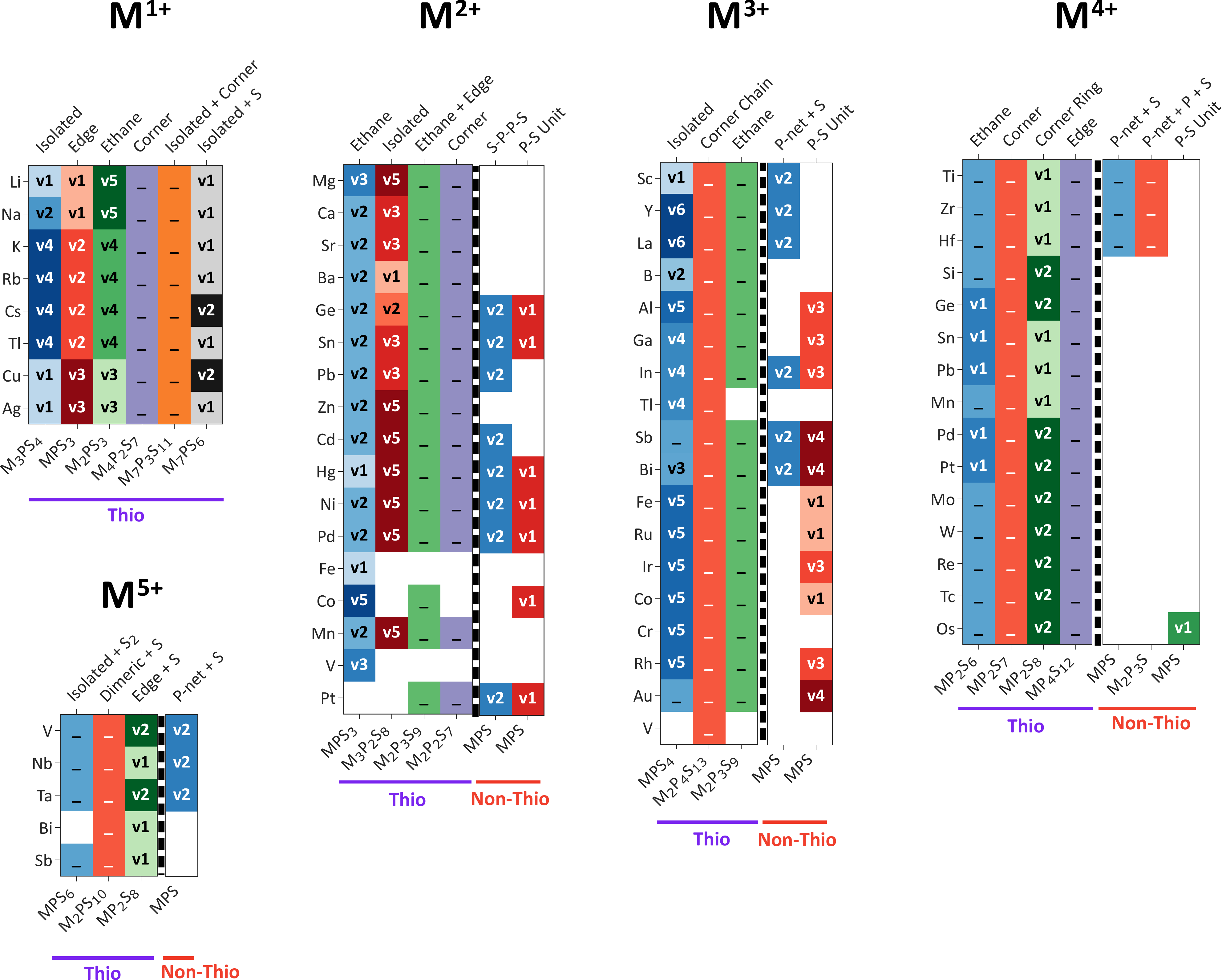}
\caption{Prototype heat maps of lowest-energy phosphosulfide polymorphs calculated in this study. Both the color and the number in each cell indicate the lowest-energy prototype for a given composition and oxidation state of the metal. The "\_" label indicates that there was only one known prototype for that combination of composition and oxidation state of the metal. White cells indicate that the corresponding material was not calculated. The visualization concept is otherwise analogous to Fig.~\ref{fig:HM_EaH} in the main article.}
\label{fig:HM_Versions}  
\end{figure*}

\subsection{Choice of elements, structural prototypes, and structure recognition}
At the time of writing, the Materials Project contained 168 ternary M$_x$P$_y$S$_z$ materials containing phosphorus and sulfur simultaneously. From these, 81 unique structural prototypes were identified. 71 of these prototypes were retained and used in our high-throughput screening. They are listed in Table~\ref{tab:mp-id} Here we list the applied filters and other choices made during the screening:

\begin{itemize}
\item We excluded prototypes where M was one of the following elements: H,  O, N, C, As, Te, the halogens, and the noble gases.
\item We excluded the elements listed above, plus Be, lanthanides, and any element with higher atomic number than Bi as the metals M that we substituted into the various prototypes. The only exception was La, the only lanthanide without valence electrons in f states, which therefore does not suffer from the poor description of f states in standard DFT. 
\item Structural prototypes found for lanthanide and actinide phosphosulfides were retained and used in calculations with metals with compatible states outside of the f-block. There are five such prototypes (Table~\ref{tab:mp-id}): MP$_4$S$_{12}$\_o4 (U), MPS\_o4 (Th), MPS\_v2\_o3 (Gd), MPS$_4$\_v3\_o3 (Lu) and MPS$_3$\_v4\_o2 (Eu). While four of them had also been observed in phosphosulfides of metals outside of the f-block, the \ce{MP$_4$S$_{12}$\_o4} prototype had only been reported for f-block metals U and Np. Interestingly, this prototype allowed us to discover the thermodynamically stable ZrP$_4$S$_{12}$ and HfP$_4$S$_{12}$, as well as five other materials within the 100~meV/atom stability threshold (Fig,~\ref{fig:HM_EaH} in the main article).
\item Only prototypes exhibited by materials with a calculated energy above hull below 200~meV/atom on Materials Project were retained.
\item We excluded prototypes that were slight variations of another existing prototype. This filter includes materials that are supercell versions of a simpler structure with very small distortions, and materials with very large stoichiometric coefficients that are compositionally and structurally very close to a simpler structure. For example, Ta$_4$P$_4$S$_{29}$ has the same structural motifs as TaPS$_6$, but with an extra one dimensional sulfur chain along the polyanionic structure.
\item Among the ternary phosphosulfides reported in Materials Project, \ce{Ga5P3S3} (mp-676038) and \ce{Y2PS} (mp-1216056) turn out to be ordered versions of materials that have only been synthesized as (disordered) solid solutions. This relatively common discrepancy between computational predictions and experimental realizations of materials has often been ignored in the past, but it is becoming evident that it needs to be addressed for effective computation-driven materials discovery.\cite{leemanChallengesHighThroughputInorganic2024a}. The synthesized version of \ce{Ga5P3S3} (ICSD collection code 108488) exhibited full P-S disorder and partial Ga occupation. The synthesized version of \ce{Y2PS} (ICSD collection code 44976) also exhibited full P-S disorder. Interestingly, the calculated ordered versions of these compounds lie on the stability hull (\ce{Y2PS}) or very close to it (6 meV/at for \ce{Ga5P3S3}), despite the fact that \ce{Y2PS} does not seem to be a charge-neutral composition. The disordered versions of these materials are likely to be even more stable. Thus, we used the two corresponding prototypes (\ce{M5P3S3}\_o3 and \ce{M2PS}\_o3) in our screening and substituted other trivalent metals to calculate more accurate convex hulls. However, we do not present the resulting data in the analysis of phosphosulfides materials and properties, since the study of phosphosulfide solid solutions goes beyond the scope of this paper.
\item After applying the filters above, very few prototypes turned out to have a structural dimensionality lower than 2D. These are \ce{Cr2P3S12}, \ce{V2PS10} and four versions of trivalent metals in the \ce{MPS4} composition. These prototypes were considered, but isovalent replacement of the metals was only performed for a few metals similar in chemistry and size to the original reported. The only two materials with less than 2D dimensionality that turned out to be the lowest-energy polymorph at their composition and be within the \SI{100}{meV/atom} stability threshold on the convex hull are \ce{AuPS4} and \ce{SbPS4}.
\item Since both the lattice parameters and the atomic positions were relaxed in the calculation, prototypes incorporating a different metal than the original would sometimes relax into a noticeably different structure. However, there were no cases where the polyanion motifs shown in Fig.~\ref{fig:motifs} were qualitatively disrupted upon relaxation. Therefore, all the lowest-energy polymorphs shown in Fig.~\ref{fig:HM_EaH} still belong to the polyanion structural categories of their original prototype. In the supplementary data accompanying this paper, we specify whether the relaxed structure of each material is exactly in the same prototype, in a distorted version of the original prototype, or in a radically different version of the original prototype.
\end{itemize}

\setlength{\LTcapwidth}{1\linewidth}
\begin{longtable}[c]{lcl}
\caption{List of the 71 structural prototypes used in the screening. The second and third column indicate the metal present in the original version of the prototype, together with its Materials Project ID. The prototype label is M$_x$P$_y$S$_z$\_v$_j$\_o$_k$, where M$_x$P$_y$S$_z$ is the general formula of the composition. v$_j$ is used to distinguish different structural prototypes for that composition as v$_1$, v$_2$, v$_3$. o$_k$ is used to the label the oxidation state $k$ of metal elements expected to yield a charge-neutral composition. \label{tab:mp-id}}
\\
\hline
\multicolumn{1}{c}{\textbf{Prototype}} & \multicolumn{1}{c}{\textbf{Ref. (Metal)}} & \multicolumn{1}{c}{\textbf{Ref. (mp-id)}} \\
\hline 
\endfirsthead
\hline
\multicolumn{1}{c}{\textbf{Prototype}} & \multicolumn{1}{c}{\textbf{Ref. (Metal)}} & \multicolumn{1}{c}{\textbf{Ref. (mp-id)}} \\ \hline
\endhead
\hline
\endlastfoot
M$_2$PS$_3$\_v1\_o1      & Li                         & mp-38200                   \\
M$_2$PS$_3$\_v2\_o1      & Li                         & mp-1222635                 \\
M$_2$PS$_3$\_v3\_o1      & Ag                         & mp-558469                  \\
M$_2$PS$_3$\_v4\_o1      & Rb                         & mp-1196612                 \\
M$_2$PS$_3$\_v5\_o1      & Ag                         & mp-561822                  \\
M$_3$PS$_4$\_v1\_o1      & Ag                         & mp-12459                   \\
M$_3$PS$_4$\_v2\_o1      & Na                         & mp-28782                   \\
M$_3$PS$_4$\_v3\_o1      & Li                         & mp-2646995                 \\
M$_3$PS$_4$\_v4\_o1      & Tl                         & mp-16848                   \\
M$_3$PS$_4$\_v5\_o1      & Li                         & mp-985583                  \\
M$_3$PS$_4$\_v6\_o1      & Li                         & mp-1097036                 \\
M$_4$P$_2$S$_7$\_o1         & Ag                         & mp-27482                   \\
M$_7$P$_3$S$_11$\_o1        & Li                         & mp-641703                  \\
M$_7$PS$_6$\_v1\_o1      & Cu                         & mp-1196216                 \\
M$_7$PS$_6$\_v2\_o1      & Li                         & mp-1211324                 \\
MPS$_3$\_v1\_o1       & Ag                         & mp-5470                    \\
MPS$_3$\_v2\_o1       & Cs                         & mp-504838                  \\
MPS$_3$\_v3\_o1       & Cu                         & mp-1105187                 \\
M$_2$P$_2$S$_7$\_o2         & Hg                         & mp-27171                   \\
M$_2$P$_3$S$_9$\_o2         & Zn                         & mp-27656                   \\
M$_3$P$_2$S$_8$\_v1\_o2     & Ba                         & mp-561443                  \\
M$_3$P$_2$S$_8$\_v2\_o2     & Ba                         & mp-554255                  \\
M$_3$P$_2$S$_8$\_v3\_o2     & Pb                         & mp-28140                   \\
M$_3$P$_2$S$_8$\_v4\_o2     & Hg                         & mp-1212489                 \\
M$_3$P$_2$S$_8$\_v5\_o2     & Zn                         & mp-30311                   \\
MP$_2$S$_6$\_v1\_o2      & Ni                         & mp-769218                  \\
MP$_2$S$_6$\_v2\_o2      & Sn                         & mp-36381                   \\
MPS\_v1\_o2        & Ir                         & mp-1220004                 \\
MPS\_v2\_o2        & Pd                         & mp-7280                    \\
MPS$_3$\_v1\_o2       & Fe                         & mp-1232490                 \\
MPS$_3$\_v2\_o2       & Sn                         & mp-13923                   \\
MPS$_3$\_v3\_o2       & Ni                         & mp-676040                  \\
MPS$_3$\_v4\_o2       & Eu                         & mp-20217                   \\
MPS$_3$\_v5\_o2       & Cd                         & mp-9330                    \\
MP$_3$S$_9$\_o3          & Cr                         & mp-675980                  \\
M$_2$P$_4$S$_{13}$\_o3        & V                          & mp-620190                  \\
MPS$_3$\_o3           & Co                         & mp-676437                  \\
M$_2$P$_3$S$_9$\_o3         & Al                         & mp-1193708                 \\
M$_2$PS\_o3           & Y                          & mp-1216056                 \\
M$_5$P$_3$S$_3$\_o3         & Ga                         & mp-676038                  \\
MPS\_v1\_o3        & Os                         & mp-1102534                 \\
MPS\_v2\_o3        & Gd                         & mp-1191365                 \\
MPS\_v3\_o3        & Ni                         & mp-505820                  \\
MPS\_v4\_o3        & Au                         & mp-1094079                 \\
MPS$_4$\_v1\_o3       & Sc                         & mp-6999                    \\
MPS$_4$\_v2\_o3       & Ga                         & mp-30979                   \\
MPS$_4$\_v3\_o3       & Lu                         & mp-30287                   \\
MPS$_4$\_v4\_o3       & In                         & mp-20790                   \\
MPS$_4$\_v5\_o3       & Cr                         & mp-542096                  \\
MPS$_4$\_v6\_o3       & Bi                         & mp-27133                   \\
MP$_2$S$_6$\_o4          & Ti                         & mp-13666                   \\
MP$_2$S$_7$\_o4          & Zr                         & mp-31014                   \\
MP$_2$S$_8$\_v1\_o4      & Nb                         & mp-28130                   \\
MP$_2$S$_8$\_v2\_o4      & Nb                         & mp-559923                  \\
MP$_4$S$_{12}$\_o4         & U                          & mp-1201721                 \\
M$_2$P$_3$S\_o4          & Zr                         & mp-1215423                 \\
MPS\_o4            & Th                         & mp-12876                   \\
MP$_2$S$_8$\_v1\_o5      & Nb                         & mp-28130                   \\
MP$_2$S$_8$\_v2\_o5      & Nb                         & mp-559923                  \\
M$_2$PS$_{10}$\_o5         & Nb                         & mp-648932                  \\
MPS\_v1\_o5        & Os                         & mp-1102534                 \\
MPS\_v2\_o5        & Nb                         & mp-27471                   \\
MPS$_6$\_o5           & Ta                         & mp-27673                   \\
M$_2$PS$_{10}$\_v2\_o5           & V                         & mp-648414                   \\
M$_2$P$_3$S$_{12}$\_o3           & Cr                         & mp-778446                   \\
MPS$_4$\_v7\_o3           & Au                         & mp-30938                   \\
MPS$_4$\_v8\_o3           & Sb                         & mp-572597                   \\
MPS$_4$\_v9\_o3 
& Al                         & mp-555538                   \\
MPS$_4$\_v10\_o3 
& Al                         & mp-27462
\\
MPS$_4$\_v11\_o3 
& Al                         & mp-1102285 
\\
MPS$_4$\_v12\_o3 
& Al                         & mp-1071955  \\
\end{longtable}

\subsection{Oxidation states}

Balancing oxidation states to generate charge-neutral compositions in phosphosulfides is often not straightforward. We found that the automatic oxidation state assignment implemented in pymatgen was often unreliable with phosphosulfide compounds, possibly due to the ambiguity in the oxidation state of phosphorus and of many metals. Therefore, we manually checked the pymatgen-suggested oxidation states for all the screened materials and manually reassigned oxidation states when they clashed with chemical intuition. Here we list our criteria for assigning oxidation states.

\begin{itemize}
\item When possible, we use the generalized $8 - N$ rule~\cite{Pearson1964} for assigning formal charges.
\item We always assume that S is in the -2 oxidation state when it doesn't form a bond with another S atom. If it does form such a bond, we assume that it is a single bond and assign an oxidation state of -1 to each S atom in a \ce{[S2]^{-2}} dimer.
\item For P atoms, we always assume that P is in the +5 oxidation state in the polyanions based on \ce{PS4} tetrahedra (where no P-P bonds are present), and in the +4 oxidation state in the ethane-like polyanions (where one P-P bond per P atom is present, see Fig.~\ref{fig:motifs}). For the few cases where the automatic method assigned a different oxidation state than in the rules above, we verified that such an assignment was likely triggered by multiple possible oxidation states of metal atoms rather than by a real change in bond lengths or coordination environment of the P atoms.
\item for P atoms, we always assume that P is in the -3 oxidation state when it only bonds to metal atoms. Although this is the standard situation in traditional phosphide semiconductors, the only case where this occurs among known phosphosulfide prototypes is for one half of the P atoms in the \ce{M2P3S} composition (see Fig.~\ref{fig:motifs}).
\item The most challenging oxidation state assignments are those for non-thiophosphates. Automatic recognition of oxidation states is least reliable within this family. There are two reasons for this. First, P is bonded to both M and S atoms (P-S unit), or to both M and other P atoms (P-net based structures), or even to all three at the same time (S-P-P-S unit). These features, shown in Fig.~\ref{fig:motifs}, make the oxidation state of phosphorus ambiguous. Second, non-thiophosphates are mainly formed with open-shell transition metals (Fig.~\ref{fig:HM_EaH}) which often have several plausible oxidation states themselves. In the bullet points below, we specifically discuss non-thiophosphates structures.
\item in the non-thiophosphate structures based on P-S units (Fig.~\ref{fig:motifs}), P is bonded to both S and M, so it could in theory exhibit any oxidation state except for its extreme values of -3 and +5. Transition metals that are normally trivalent make up the largest fraction of MPS compounds with P-S units. These compounds exhibit non-zero band gaps (Fig.~\ref{fig:HM_BG}), indicating that the overall formal charge on the PS unit is expected to be -3. Thus, P must be present with an oxidation state of -1 assuming the usual oxidation state of -2 for S. OsPS is the only example of a non-thiophosphate with P-S units that is much more likely to be tetravalent than trivalent. It is located on the stability hull, has been synthesized, and has a non-zero band gap (Fig.~\ref{fig:HM_BG}). Hence, we conclude that the P-S unit can also have a formal charge of -4, which implies another allowed P oxidation state of -2 for this motif. Finally, MPS materials with P-S units can also exhibit relatively low energies above hull with normally divalent transition metals (Ni and Pd), but both materials have metallic band structures. Hence, we assume that the P oxidation state of zero necessary to achieve a charge-neutral composition with divalent metals is too high and postulate that these compounds are intrinsically metallic.

\begin{table}[b!]
\begin{tabular}{llll}
\multicolumn{1}{l}{} &  & \textbf{Structural} &  \textbf{Machine-learned} \\
\multicolumn{1}{l}{\textbf{Material}} & \textbf{Prototype} & \textbf{dimensionality} & \textbf{band gap (eV)}  \\ \hline
\ce{Al2P4S13}   & \ce{M2P4S13}\_o3       & 2D & 3.15 \\
\ce{Bi2P4S13}   & \ce{M2P4S13}\_o3       & 2D & 2.57 \\
\ce{Cs2PS3}     & \ce{M2PS3}\_v4\_o1     & 3D & 3.46 \\
\ce{Cs3PS4}     & \ce{M3PS4}\_v4\_o1     & 3D & 3.48 \\
\ce{GePS3}      & \ce{MPS3}\_v2\_o2      & 2D & 2.78 \\
\ce{HfP2S8}     & \ce{MP2S8}\_v1\_o4     & 3D & 2.71 \\
\ce{HfP2S7}     & \ce{MP2S7}\_o4         & 3D & 2.70 \\
\ce{HfP4S12}    & \ce{MP4S12}\_o4        & 3D & 2.77 \\
\ce{Hf2P3S}     & \ce{M2P3S}\_o4         & 3D & 0.01 \\
\ce{Sb2P4S13}   & \ce{M2P4S13}\_o3       & 2D & 2.61 \\
\ce{Sc2P4S13}   & \ce{M2P4S13}\_o3       & 2D & 2.98 \\
\ce{ScPS}       & \ce{MPS}\_v2\_o3       & 3D & 1.15 \\
\ce{Sr3P2S8}    & \ce{M3P2S8}\_v3\_o2    & 3D & 3.28 \\
\ce{TiP2S8}     & \ce{MP2S8}\_v1\_o4     & 3D & 1.75 \\
\ce{VP2S8}      & \ce{MP2S8}\_v2\_o5     & 2D & 1.19 \\
\ce{ZrPS}       & \ce{MPS}\_o4           & 3D & 0.17 \\
\ce{ZrP2S8}     & \ce{MP2S8}\_v1\_o4     & 3D & 2.49 \\
\ce{ZrP4S12}    & \ce{MP4S12}\_o4        & 3D & 2.62 \\
\ce{SiP2S6}     & \ce{MP2S6}\_o4         & 3D & 2.87 \\
\hline
\end{tabular}
\caption{Novel materials discovered in this work which lie on the stability hull ($E_h = 0$). The materials are based on known structural prototypes but have compositions that were unreported in both Materials Project and ICSD. \label{tab:list_new_materials}} 
\end{table}

\item for the non-thiophosphate structures based on S-P-P-S units (Fig.~\ref{fig:motifs}) P is bonded to S, to M, and to another P atom. We may expect a slightly higher P oxidation state compared to the P-S unit, due to the additional P-P bond. Indeed, these types of non-thiophosphates tend to be stable with the usually divalent metals Hg, Pd, and Pt, implying a P oxidation state of 0. HgPS, PdPS, and PtPS are confirmed to be charge balanced by calculation of a zon-zero band gap for all of them (Fig.~\ref{fig:HM_BG}).
\item for the structures that incorporate P nets (Fig.~\ref{fig:motifs}), each phosphorus atom appears to be bonded to three other phosphorus atoms and one metal atom assuming typical bond lengths. Thus, we assign an oxidation state of -1 to all phosphorus atoms within the net. This implies that \ce{MPS} compositions with the P-net + S motif are not charge balanced in conjunction with non-trivalent metals (Fig.~\ref{fig:motifs}). Therefore, we expect TiPS, ZrPS, HfPS, NbPS, and TaPS to be intrinsically metallic, but we expect ScPS, YPS, and LaPS to be (potential) semiconductors. The expectation is confirmed by our band gap calculations (Fig.~\ref{fig:HM_BG})
\item When choosing metals with oxidation states expected to give charge-neutral compositions, we made no attempt to be comprehensive with the many possible oxidation states of open-shell transition metals, and limited our screening to a maximum of two oxidation states for each metal. This is the reason for the empty cells in Fig.~\ref{fig:HM_EaH}.
\end{itemize}

\begin{figure}[ht!]
\centering%
\includegraphics[trim= 0 0 0 30, clip, width=0.6\textwidth]{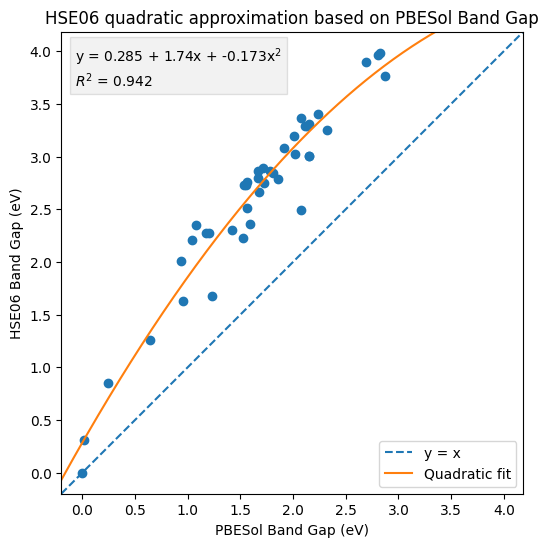}
\caption{Phosphosulfide-specific second-order polynomial regression between HSE and PBEsol band gaps. The data points are 39 ternary phosphosulfides for which HSE band gaps are available on the SNUMAT database~\cite{Kim2020b}. The PBEsol band gaps are the ones calculated in this work. This approach is analogous to the HSE to PBE regression method employed in previous work~\cite{mittmannPhosphosulfideSemiconductorsOptoelectronics2024}. The best-fit parameters and fit quality are indicated. Note that the second-order polynomial fit would become unphysical for materials with PBEsol band gaps above about approximately 3~eV.}
\label{fig:HSE Fit}
\end{figure}

\subsection{Multi-fidelity machine learning}
Here, we give further details about the architecture of the base and translation (sub)models that make up the multi-fidelity machine-learning translation pipeline used in this work. 
We recall the context in which we present these models: we want to map the PBEsol-calculated band gaps to their estimated experimental equivalents. Before discussing the specifics of the models, we summarize the structure of the so-called translation pipeline. First, we use a trained transformer-based multi-fidelity band gap predictor, referred to as the \textit{base model}, to generate two learned vector representations of the compound at two fidelity levels: the target fidelity (e.g., experimental) and the source fidelity (e.g., PBEsol). Second, the two vector representations of the compounds and the original band gap value are fed into a simple regressor MLP layer (the \textit{translation} layer/submodel) to estimate the band gap at the target fidelity.\\

\begin{figure*}[t!]
\centering%
\includegraphics[trim= 0 150 0 0, clip, width=0.8\textwidth]{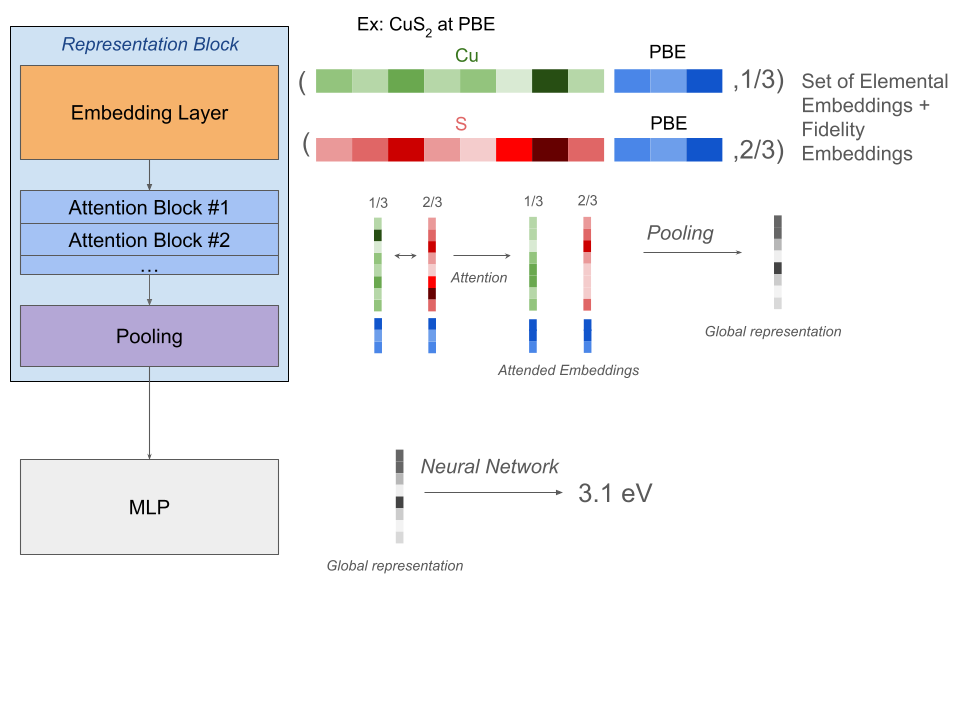}
\caption{Base model diagram depicting inference pipeline of a single entry (eg; CuS$_2$). The embedding layer produces an embedding for each element: a concatenation of a specific elemental embedding and a shared fidelity embedding. The permutation invariant transformer produces refined embeddings (residual connections not depicted). The transformation from elemental embeddings and a set of fractional weights to a global compound representation is performed by one of several possible pooling mechanisms (treated as a categorical hyperparameter). Finally, an MLP predicts the band gap from this global representation vector.}
\label{fig:ML_model_diagram}
\end{figure*}

\begin{figure*}[b!]
\centering%
\includegraphics[trim= 0 70 0 0, clip, width=0.8\textwidth]{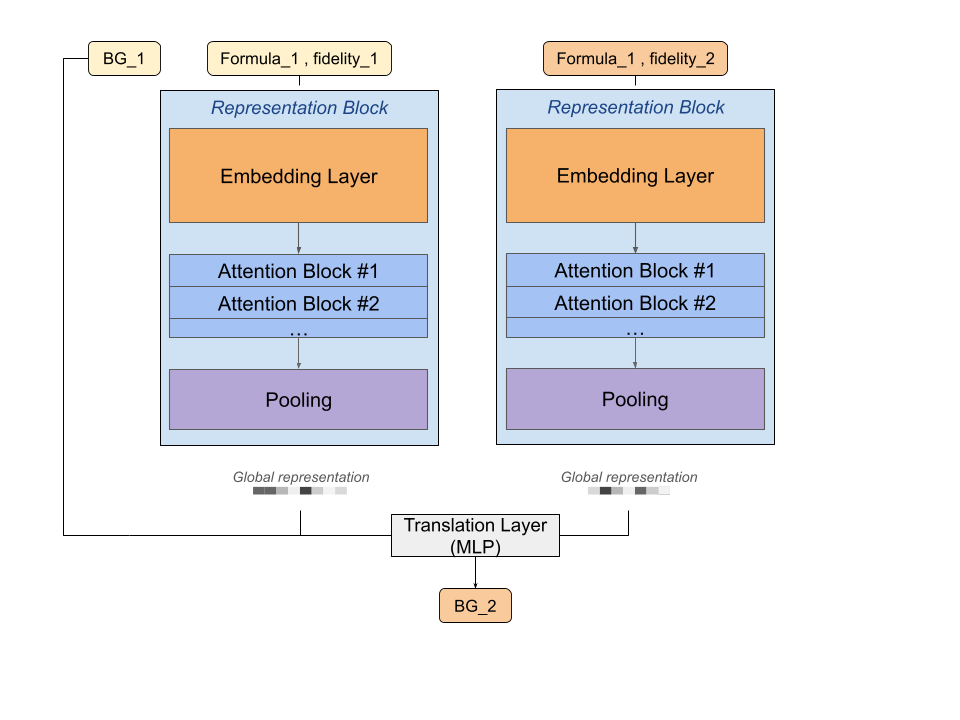}
\caption{Translation pipeline diagram. The formula and the two fidelity labels, with fidelity\_1 representing the source and fidelity\_2 representing the target fidelity, are used to generate two global representation vectors using the trained \textit{representation block} from the trained base model. The source band gap and the two representation vectors are fed into the translation layer which is just a MLP regressor layer. The output is the desired band gap at the target fidelity level. }
\label{fig:ML_translation}
\end{figure*}

\textbf{Base Model:} The base model first encodes the chemical composition and fidelity level into a set of elemental fractions and learned embeddings. An attention step is performed by a permutation invariant transformer to produce a set of refined embeddings and a pooling mechanism then aggregates these features into a single vector, the global representation, which describes the compound at that fidelity level~\cite{Wang2021b}. Finally, this global representation is fed into a multi-layer perceptron (MLP) to predict the band gap. We trained this model using entries containing a chemical composition, a band gap, and a fidelity label (indicating the experimental fidelity level or one of the six DFT fidelity levels). A diagram of this architecture is shown in Fig.~\ref{fig:ML_model_diagram}. 

\textbf{Translation layer:} The starting point of the translation layer is a fully trained multi-fidelity base model. We replaced the final MLP layer of the base model with a new, task-specific "translation layer". For the case of a PBEsol to experimental band gap translation, this new layer takes three inputs: (i) The global representation of the material at the known (PBEsol) fidelity level (generated by the base model); (ii) The global representation of the material at the target (experimental) fidelity level (also from the base model); (iii) The band gap value at the known fidelity level. The architecture of this translation model is depicted in Fig.~\ref{fig:ML_translation}.

\begin{figure}[b!]
    \centering
    \includegraphics[width=0.6\linewidth]{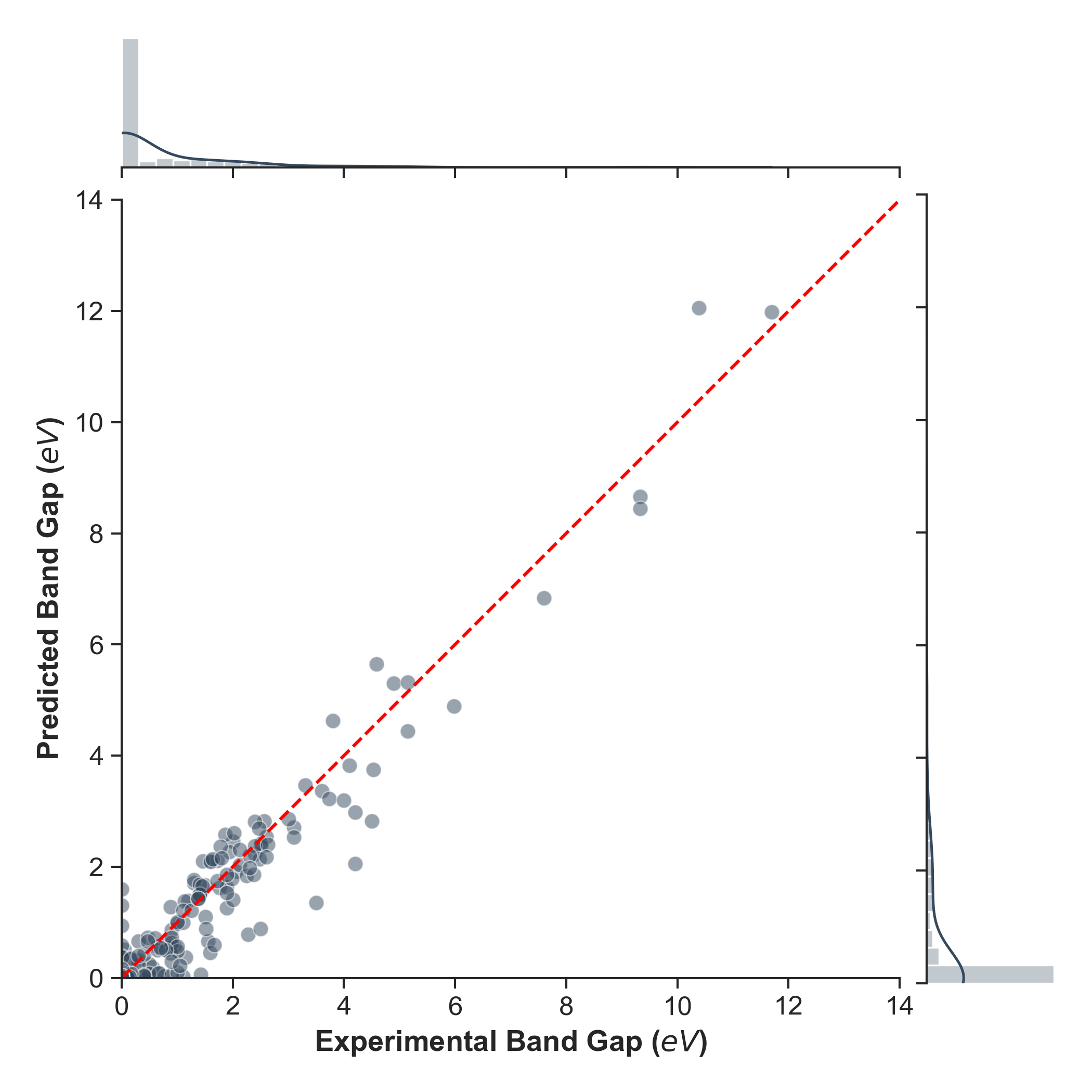}
    \caption{Predicted experimental band gap (from the PBEsol-to-experiment translation model) versus actual experimental band gap. The plotted data points are those in held-out test dataset. MAE$=0.168$, $\text{R}^2=0.876$ }
    \label{fig:pbesol_to_expt}
\end{figure}

\textbf{Full translation pipeline}. The base model, originally developed for the task of predicting band gaps from composition at different levels of DFT theory and experiment, allows for the assignment of input data to user-defined categories -- the fidelity levels.  In the context of this paper, we collected and processed freely available data into 7 fidelity categories : PBE, PBE+U, PBESol, SCAN, GLLB-SC, HSE and experimental (see Methods section in the main article). This dataset was further augmented with our in-house PBEsol calculations on ternary phosphosulfides.  
We conducted small-scale hyper-parameter optimization using the Optuna framework \cite{akibaOptunaNextgenerationHyperparameter2019}, completing approximately 130 trials. A trial's performance was evaluated against a held-out HSE test set. HSE was selected as the target trial metric because it strikes a balance band gap accuracy and data availability. Our optimization identified the categorical pooling mechanism, discrete elemental embedding dimension, and the number of attention blocks as the most critical hyper-parameters. Table~\ref{tab:base} summarizes the base model's performance on the task of predicting band gaps of unseen materials from composition.  
The base model’s good performance (summarized in Table \ref{tab:base}) demonstrates its capacity to extract robust, fidelity-aware features from chemical compositions. A reasonably low prediction error across diverse fidelity levels indicates that the internal global representation vectors effectively capture the underlying physical trends of the materials. By utilizing these pre-trained, high-quality embeddings as fixed feature extractors, the translation layer can more effectively map the latent relationship between PBEsol and experimental data. This two-stage approach ensures that the translation model benefits from a rich chemical context, even when experimental training data is relatively sparse. The base model's representation vectors for the compound at the source and target fidelity levels as well as the source band gap value are fed into the translation layer as described above.

\begin{table}[t!]
\centering
\begin{tabular}{llll}
\toprule
\textbf{Fidelity} & \textbf{MAE} \hspace*{4mm}  & \textbf{RMSE} \hspace*{4mm} & \textbf{R$^2$} \\
\textbf{level} & \textbf{(eV)}&  \textbf{(eV)} &  \\
\midrule
PBEsol& 0.0899 & 0.2377 & 0.9504 \\
SCAN     & 0.1171 & 0.2948 & 0.9409 \\
GLLB-SC  & 0.4065 & 0.5380 & 0.9279 \\
HSE     & 0.4787 & 0.6686 & 0.8223 \\
Experimental     & 0.3705 & 0.7469 & 0.7541 \\
\bottomrule
\end{tabular}
\caption{Performance metrics (mean absolute error, root mean square error, and $R^2$) for the base models predicting band gaps of different DFT functional, and predicting experimental band gaps. The errors are based on a held-out test set. \label{tab:base}}
\end{table}

\begin{table}[t!]
\centering
\begin{tabular}{llllll}
\toprule
\textbf{Source} \hspace*{4mm} & \textbf{Target}  & \textbf{MAE} & \textbf{RMSE}   & \textbf{R$^2$}  & \textbf{Test set} \\
\textbf{Fidelity} & \textbf{Fidelity}  & \textbf{(eV)} & \textbf{(eV)} & & \textbf{size} \\
\midrule
PBEsol & Experimental \hspace*{4mm} & $0.168 \pm 0.014$ \hspace*{4mm} & $0.361 \pm 0.083$  \hspace*{4mm} & $0.876 \pm 0.127$ \hspace*{4mm} & 479 \\
PBEsol & GLLB-SC     & $0.252 \pm 0.025$ & $0.358 \pm 0.052$ & $0.961 \pm 0.014$ & 281 \\
PBEsol & SCAN        & $0.063 \pm 0.015$ & $0.124 \pm 0.012$ & $0.990 \pm 0.002$ & 112,723 \\
\bottomrule
\end{tabular}
\caption{Performance metrics for the translation submodel mapping from PBEsol to various target fidelities. Results represent the mean and standard deviation across 5-fold cross-validation. \label{tab:trsl}}
\end{table}

To train the translation layer, we constructed a dataset of materials with band gap values available across multiple fidelity levels. To ensure physical consistency, we restricted the data to the lowest-energy polymorph for each composition, using Materials Project IDs (mp-ids) to verify structural correspondence whenever available.
The translation model was trained on a set of available cross-fidelity pairs in the data used for the base model, including inter-functional translations (e.g., PBE to SCAN) in addition to the PBEsol-to-experimental target. Table \ref{tab:trsl} summarizes the 5-fold cross-validation (CV) performance of the translation model on the task of translating PBEsol band gaps to a band gap at another fidelity level. Figure \ref{fig:pbesol_to_expt} shows the compiled predictions of the experimental band gap translated from the PBESol band gap over the 5 CV folds.

\section{Extended experimental details}

The combinatorial DADMARS thin film synthesis approach was realized in a custom reactive sputter chamber (Kurt J. Lesker) equipped with magnetron sputtering sources for metallic species (here, Cu and Ag), a sulfur cracker with three heating zones (Nano4Energy/Gencoa), and separate gas inlets for the \ce{PH3} reactive gas and the Ar working gas. Combinatorial gradients in thin film composition are obtained by suitable orientation of the magnetron sources and the sulfur cracker injection tube with respect to a stationary substrate, as described in Ref.~\cite{mittmannLargeareaThinfilmSynthesis2025a}. The metals fluxes are tuned by adjusting the RF power applied to the sputter targets with respect to ground. The sulfur flux is tuned by adjusting the duty cycle of a pulsed valve. The \ce{PH3} flux is tuned by adjusting its flow rate through a mass flow controller. At a total synthesis pressure of \SI{0.67}{Pa}, typical deposition rates and overall partial pressures of sulfur and \ce{PH3} are \SI{1}{Å/s}, \SI{0.05}{Pa}, and \SI{0.02}{Pa}, respectively. A typical base pressure of the chamber is \SI{2e-5}{Pa}. The films were deposited on a combination of crystalline silicon and soda lime glass substrates, covering a total area of about $(10 \times 8)$~cm$^2$. The sputter chamber is directly connected to an actively purified glovebox filled with \ce{N2} (LC Technology) so the thin films and sputter targets can be retrieved and stored without air exposure. Nevertheless, the four thin film materials synthesized in this work all appeared to be stable in air, within an observation time frame of weeks.

Elemental composition was automatically mapped as a function of position throughout the combinatorial samples (typically $10 \times 10$ measurement points for each combinatorial deposition) by energy dispersive X-ray spectroscopy (EDX). The measurements were conducted in a FEI Quanta FEG 250 scanning electron microscope equipped with an Oxford Instruments EDX detector. The LayerProbe analysis package (Oxford Instruments) was used to deconvolve the film and substrate contributions to the spectra and obtain reliable thin-film composition results with a high throughput~\cite{mittmannLargeareaThinfilmSynthesis2025a}, resulting in the plots shown in Fig.~\ref{fig:ternary_systems}.

To determine if a given composition gave rise to any of the computationally screened crystal structures, we performed X-ray diffraction (XRD) measurements in the parallel beam $\theta$-$2\theta$ geometry across the combinatorial libraries (typically $10 \times 10$ measurement points for each combinatorial deposition) with a Rigaku SmartLab instrument. Employing an automated mapping stage, a high-power rotating Cu-K$\alpha$ source, focusing optics, and a 2D detector enabled a high measurement throughput. The background was subtracted with a b-spline function. The expected angles of the XRD reflections for the synthesized materials were calculated from the crystal structures available on the ICSD database using a dedicated calculator in VESTA~\cite{Momma2011}.

\begin{figure}[t!]
\centering%
\includegraphics[width=0.63\textwidth]{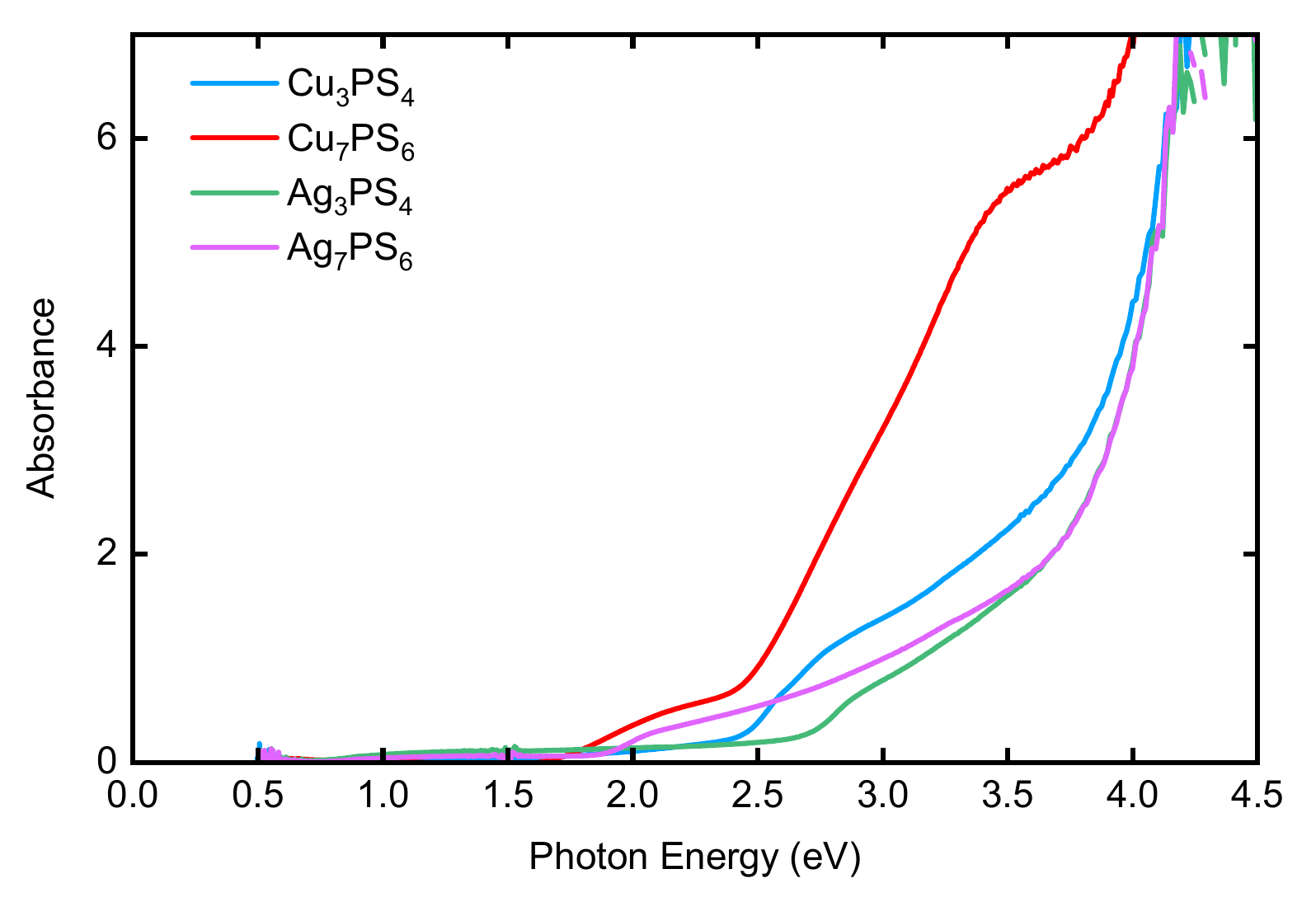}
\caption{Experimentally measured absorbance spectra $A(h\nu)$ of the four synthesized thin-film materials as a function of photon energy $h\nu$.}
\label{fig:abs}
\end{figure}

\begin{figure}[t!]
\centering%
\includegraphics[width=0.8\textwidth]{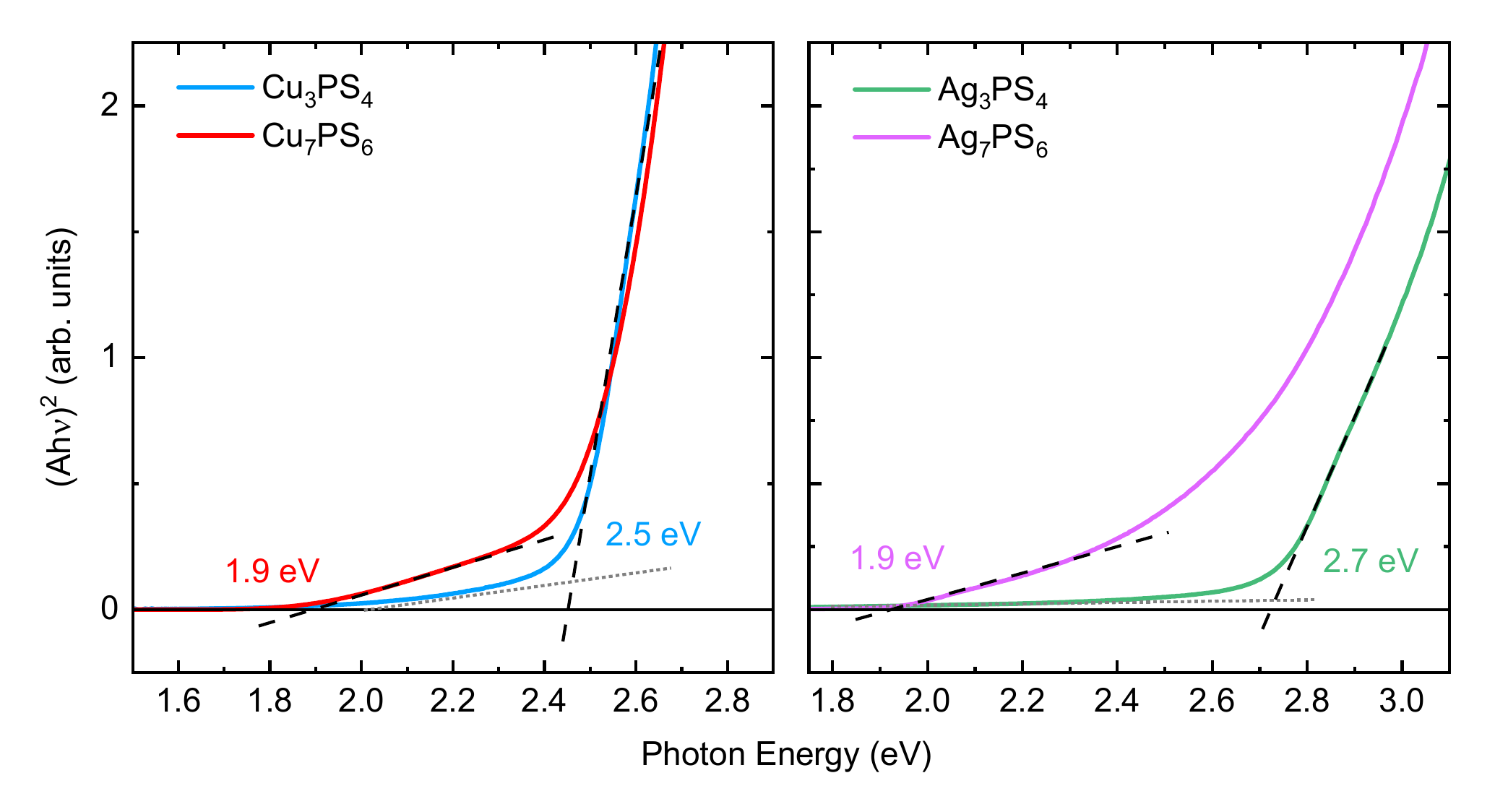}
\caption{Tauc plots for direct or slightly indirect gap materials ($(A h\nu)^2$ versus $h\nu$) applied to the four synthesized thin-films. The photon energy of the absorbance onset is extrapolated by fitting the near-band-gap absorbance with a straight (dashed) line and finding its intercept with background absorption (dotted line). The band gap values are indicated.}
\label{fig:tauc}
\end{figure}

For the individual data points that resulted in single-phase materials with a well defined crystal structure, we determined band gaps by measuring the direct and diffuse components of their optical transmission $T(h\nu)$ and reflection $R(h\nu)$ spectra with a broad-band light source, an integrating sphere, and a spectrophotometer (Agilent Cary 7000). $h\nu$ is the photon energy. To derive the band gap, we first deduced the film's absorbance spectrum $A(h\nu)$ as $A = -\ln[T/(1-R)]$ (Fig.~\ref{fig:abs}). Then, we extrapolated the photon energy of the absorbance onset with a Tauc plot, by fitting the near-band-gap absorbance with a straight line in a $(A h\nu)^2$ versus $h\nu$ plot (Fig.~\ref{fig:tauc}). We estimate that the uncertainty in the experimental determination of the band gap in these thin film samples with the present measurement technique is $\pm \SI{0.1}{eV}$. The two main sources of error are (i) the often questionable applicability of the Tauc method for a generic semiconductor~\cite{Dolgonos2016a}, and (ii) the difficulty in distinguishing above-band gap absorption (especially from indirect transitions) from sub-band gap absorption due to, e.g., disorder and defects. For the case of \ce{Cu3PS4}, more detailed analysis presented in previous work~\cite{mittmannLargeareaThinfilmSynthesis2025a} revealed that an alternative band gap estimation method based on alignment of experimental and calculated absorption spectra yielded a band gap that was \SI{0.1}{eV} lower than using the Tauc plot method.

\begin{figure}[b!]
    \centering
\includegraphics[width=0.75\linewidth]{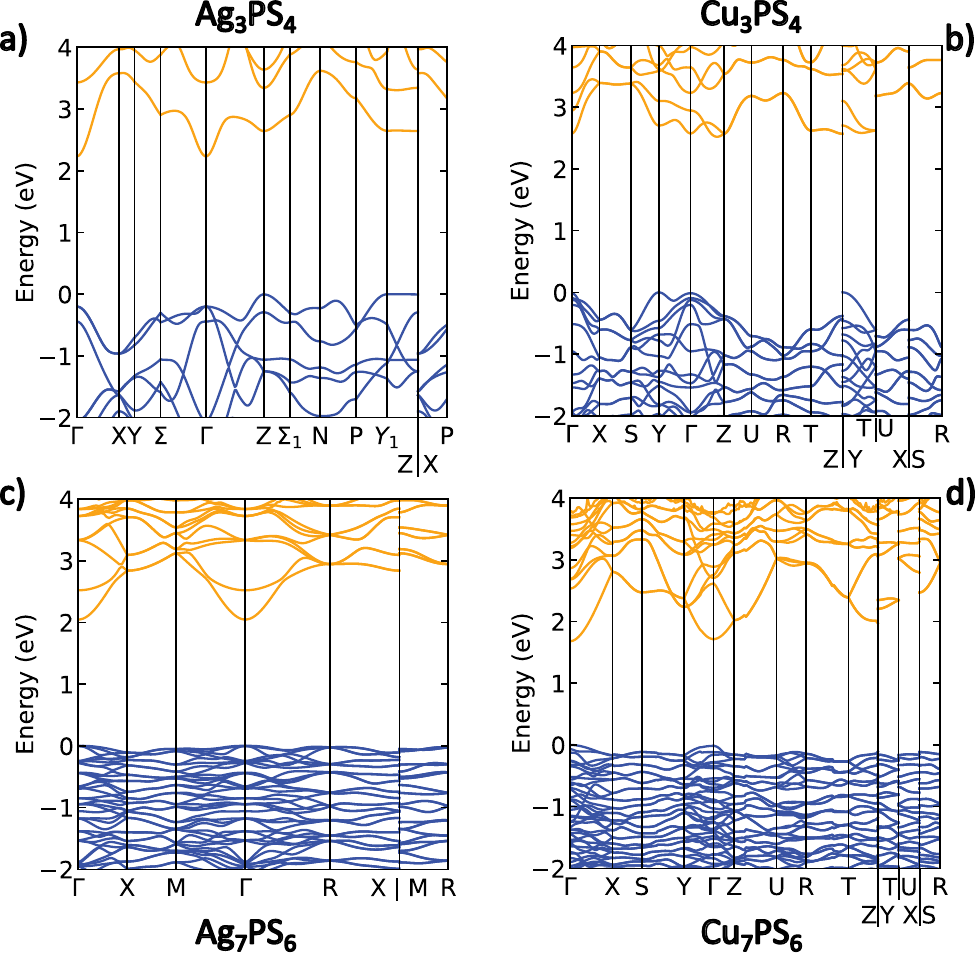}
    \caption{HSE06 band structures of the four materials synthesized as thin films: a) \ce{Ag3PS4},  b) \ce{Cu3PS4},  c) \ce{Ag7PS6} and  d) \ce{Cu7PS6}, plotted with sumo~\cite{ganoseSumoCommandlineTools2018}. The band gaps extracted from these band structures are used in Table \ref{tab:bandgaps} in the main article.}
    \label{fig:HSE_Bands}
\end{figure}

\begin{figure}[ht!]
\centering%
\includegraphics[width=0.9\textwidth]{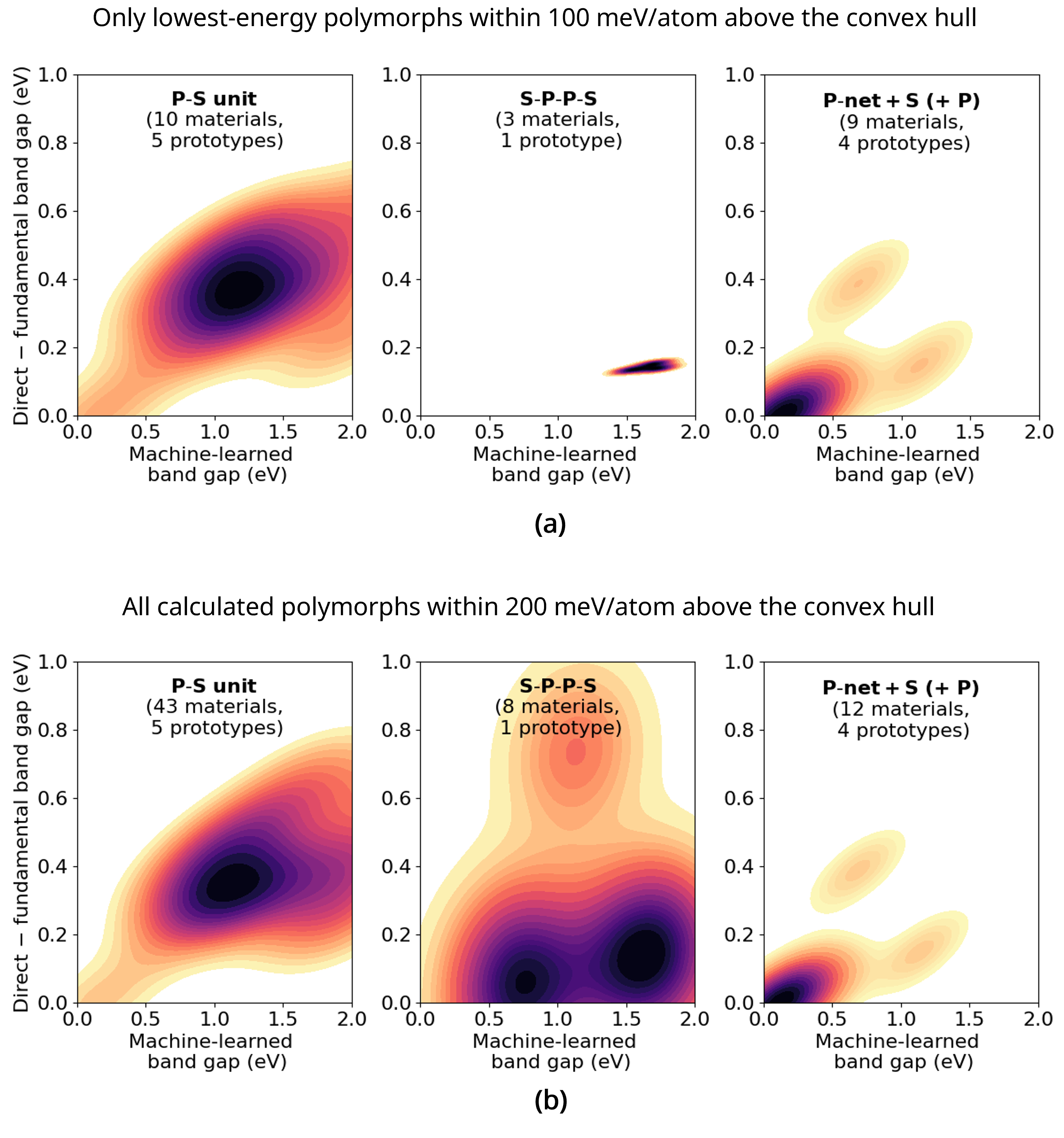}
\caption{Band gap contour plots analogous to the ones in Fig.~\ref{fig:BG_vs_indirectedness} in the main article, but only for non-thiophosphates based on the P-S unit motif (left), the S-P-P-S motif (middle), and the P-net +S based motifs (right), both with and without separated P anions. The number of materials used to construct the plots, and the number of structural prototypes considered for each structural motif, are indicated in each plot. (a): Only lowest-energy polymorphs with less than \SI{100}{meV/atom} $E_h$ are considered. (b): All polymorphs with less than \SI{200}{meV/atom} $E_h$ are considered.}
\label{fig:nonthio_BG}
\end{figure}

\end{document}